\documentclass{aa}  

\usepackage{graphicx}
\usepackage{txfonts}

\usepackage{bm}
\usepackage{upgreek}

\newcommand{\lin}[0]{_{\mathrm{in}}}
\newcommand{\lout}[0]{_{\mathrm{out}}}
\newcommand{\dd}{\,\mathrm{d}}
\newcommand{\brac}[1]{\left\langle #1 \right\rangle}
\newcommand{\vrac}[1]{\left\vert #1 \right\vert}
\newcommand{\std}[1]{\mathrm{std}\left(#1\right)}

\graphicspath{ {Pics/} }

\begin{document} 

\title{Spiral structures in gravito-turbulent gaseous disks}


\author{William B\'ethune\inst{1} \and Henrik Latter\inst{2} \and Wilhelm Kley\inst{1}}

\institute{Institut f\"ur Astronomie und Astrophysik, Universit\"at T\"ubingen, Auf der Morgenstelle 10, 72076 T\"ubingen, Germany
  \and
  DAMTP, University of Cambridge, CMS, Wilberforce Road, Cambridge, CB3 0WA, UK\\
  \email{william.bethune@uni-tuebingen.de, hl278@cam.ac.uk, wilhelm.kley@uni-tuebingen.de}
}

\date{Received ****; accepted ****}

\abstract
    {Gravitational instabilities can drive small-scale turbulence and large-scale spiral arms in massive gaseous disks under conditions of slow radiative cooling. These motions affect the observed disk morphology, its mass accretion rate and variability, and could control the process of planet formation via dust grain concentration, processing, and collisional fragmentation.}
{We study gravito-turbulence and its associated spiral structure in thin gaseous disks subject to a prescribed cooling law. We characterize the morphology, coherence, and propagation of the spirals and examine when the flow deviates from viscous disk models.}
{We used the finite-volume code \textsc{Pluto} to integrate the equations of self-gravitating hydrodynamics in three-dimensional spherical geometry. The gas was cooled over longer-than-orbital timescales to trigger the gravitational instability and sustain turbulence. We ran models for various disk masses and cooling rates.}
{In all cases considered, the turbulent gravitational stress transports angular momentum outward at a rate compatible with viscous disk theory. The dissipation of orbital energy happens via shocks in spiral density wakes, heating the disk back to a marginally stable thermal equilibrium. These wakes drive vertical motions and contribute to mix material from the disk with its corona. They are formed and destroyed intermittently, and they nearly corotate with the gas at every radius. As a consequence, large-scale spiral arms exhibit no long-term global coherence, and energy thermalization is an essentially local process.}
{In the absence of radial substructures or tidal forcing, and provided a local cooling law, gravito-turbulence reduces to a local phenomenon in thin gaseous disks.}


\maketitle

\section{Introduction}

When sufficiently massive, the dynamics of astrophysical disks is controlled by their own gravity in addition to the object they orbit. A number of observed class 0/I protostellar disks could fit in this category and thus exhibit self-gravitating phenomena \citep[e.g.,][]{eisner05,ALMA15,mann15,liu16,forgan16}. These phenomena can be global as they result from the long-range interaction of distant disk parcels. Even so, they may be described in terms of the local properties near a given disk parcel, such as the stability of the disk to gravitational collapse \citep{safronov60,toomre64,goldreich65a}. In this paper, we consider gravitationally unstable gaseous disks whose mass is a significant fraction of that of the central object. These include the outer regions of circumstellar disks or active galactic nuclei \citep{shlosman87,durisen07,kratter16}, but this excludes weakly collisional systems such as galactic disks or planetary rings \citep[e.g.,][]{salo92,richardson94}.

Gravitational instability (GI) can trigger the fragmentation of the disk into stellar and substellar companions \citep{kolykhalov80,podolak93,boss98,boss00}, provided that radiative cooling is fast enough \citep{gammie01,rice03,lin16,booth19}. We consider cooling times longer than orbital times for which the instability sustains gravito-turbulence instead. The gas orbital energy is converted into heat by the dissipation of supersonic turbulence, balancing radiative cooling while allowing matter to accrete onto the central object \citep{paczynski78,linpringle87}. Gravito-turbulence appears to organize itself into large-scale spiral arms amidst small-scale fluctuations \citep{laughlin94,riols17}. These spiral arms are believed to induce variability in the mass accretion rate \citep{lodato05,vorobyov05,vorobyov15}, and they could play important roles in planet formation via dust grain accumulation \citep{rice04,clarke09,dipierro15}, thermal processing, and collisional fragmentation \citep{boss05,boley08,ida08,booth16}. The morphology of observed spirals may also be used to identify self-gravitating disks and to constrain their mass and temperature distributions \citep{dong15,dong16,dong18,hall16,meru17,forgan18,cadman20}.

The ``shearing box'' approximation \citep{hill78,goldreich65b} has often been used to study the saturation of gravitational disk instabilities at high spatial resolution in a controlled environment \citep{gammie01,shi14,riols17,hirose17,baehr17,booth19}. However, massive disks may also support global wave modes which the shearing box cannot capture \citep{adams89,papaloizou91}. Although tightly wound waves of short wavelengths can fit in a shearing box, the assumed periodicity forces them to interact with the same fluid elements without ever leaving the box. Should a global wave train travel through the disk, its coherence would be tied to keeping a well-defined angular pattern speed \citep{lin64}. In this case, the wave could extract orbital energy at one location and transport it radially before thermalization, in contrast to the local energy conversion of viscous $\alpha$ disk models \citep{shakura73,balbuspap99} and of local boxes. As a consequence, the angular momentum flux and the resulting mass accretion rate would no longer be determined by a condition of local thermal balance \citep{gammie01}.

Several authors have addressed the issue of nonlocal energy transport in global simulations of gravito-turbulent disks. Using cylindrical grid-based simulations and focusing on low-order spirals, \citet{mejia05}, \citet{boley06b}, and \citet{steiman13} identified traveling waves by their distinct pattern speed. However, these waves could only travel over relatively narrow radial intervals. In addition, their nonlinear coupling with smaller-scale modes may have led to local energy dissipation and interfered with wave-induced transport if present \citep{michael12}. Using smoothed particle hydrodynamics (SPH), \citet{lodato04,lodato05} measured a stress compatible with local energy conversion in disks as massive as the central object. Later, \citet{cossins09} and \citet{forgan11} argued for the preponderance of nonlocal energy transport for disk mass approaching the central object's, which is in apparent conflict with \citet{lodato05}. However, their simulation diagnostics relied on the linear dispersion relation for tightly wound two-dimensional (2D) isothermal density waves, with a simple correction for the disk stratification. The use of such a dispersion relation is questionable given the nonlinear nature of these waves \citep{shu85,laughlin97,laughlin98} and the nonisothermal structure of the disk \citep{lubow98}. 

In this paper, we study the spiral structures emerging in global three-dimensional (3D) simulations of self-gravitating disks. In particular, we examine their morphology, coherence, and propagation through the disk while paying special attention to their deviations from locality. Unlike most previously published results, we used a finite-volume code in spherical geometry to capture discontinuities and to resolve the vertical motions of the flow from the disk midplane to its low-density corona. In Sect. \ref{sec:method} we present our disk model, provide some theoretical expectations, and explain the numerical method designed to confront them. We chose a simulation as a reference to present in Sect. \ref{sec:refsim} before considering different disk masses and cooling times in Sect. \ref{sec:sample}. We summarize our findings in Sect. \ref{sec:summary} and enumerate several issues left open by this work. 

\section{Model, theory, and method} \label{sec:method}

Our goal is to characterize steady gravito-turbulent disks within a global framework. In this section we describe our physical modeling of the problem, some background theory including our a priori expectations, and the numerical method we used. 

\subsection{Disk model} \label{sec:model}

We consider a rotationally supported gaseous disk orbiting a central mass $m_{\star}$. We assume that the disk is geometrically thin and that its mass $m_{\mathrm{disk}}$ is sufficiently large for its own gravity to be dynamically important. Finally, we assume that the disk radiates heat away at a prescribed rate, thereby evolving toward a cold and gravitationally unstable state. 

We place the central object at the origin of the spherical coordinate system $\left(r,\theta,\varphi\right)$, with $\theta=0$ along the rotation axis of the disk. Let also $\left(R,\varphi,z \right)$ denote cylindrical coordinates, with $z=r \cos\left(\theta\right)$ measuring the height relative to the disk midplane. We isolate the disk by considering only a finite radial interval $r \in \left[r\lin,r\lout\right]$, excluding both the central mass and the environment of the disk on larger scales. 

\subsection{Governing equations} \label{sec:equations}

The gas density $\rho$, velocity $\bm{v}$, and pressure $P$ evolve according to
\begin{align}
  \partial_t \rho &= - \nabla\cdot\left(\rho \bm{v}\right), \label{eqn:dtrho}\\
  \rho \partial_t \bm{v} &= -\rho \bm{v}\cdot\nabla \bm{v} - \nabla P - \rho\nabla\Phi, \label{eqn:dtv}\\
  \partial_t P &= -\bm{v}\cdot\nabla P - \gamma P \nabla\cdot \bm{v} - \left(\gamma-1\right) q^-, \label{eqn:dtP}
\end{align}
assuming an ideal equation of state $P=\left(\gamma-1\right)\rho \epsilon$ with $\epsilon$ the specific internal energy and $\gamma=5/3$ a constant adiabatic exponent. The gravitational potential $\Phi$ is the sum of two contributions: $\Phi_{\star}=-Gm_{\star}/r$ from the central mass and $\Phi_{\mathrm{disk}}$ from the disk itself, which satisfies Poisson's equation
\begin{equation} \label{eqn:poisson}
\Delta \Phi_{\mathrm{disk}} = 4\uppi G \rho, 
\end{equation}
and is assumed to vanish at infinity.

We took the reference frame to be inertial, hence we omitted the net force induced by the disk on the central object. Doing so precludes the development of a strong dipolar moment of the gas distribution, as manifested by eccentric or one-armed spirals \citep{adams89,heemskerk92}. One motivation for taking this approximation issues from the mass and momentum flows through the boundaries of the domain, which make it impossible to track the position of a true barycenter over time. In Appendix \ref{app:inertia} we confirm that this acceleration is unimportant as far as our diagnostics on the disk dynamics are concerned \citep[see also][]{michael10}. 

The rightmost term in the internal energy equation \eqref{eqn:dtP} represents radiative cooling, which is modeled as a local heat sink:
\begin{equation} \label{eqn:cool}
  \left(\gamma-1\right) q^- = \frac{P-\rho c^2_{\mathrm{floor}}\left(R\right)}{\beta \Omega_{\star}^{-1}}, 
\end{equation}
where $\beta$ is a prescribed constant, $\Omega_{\star} = \sqrt{G m_{\star} / R^3}$ is the local Keplerian frequency, and $c_{\mathrm{floor}}\left(R\right) \equiv 10^{-3} \Omega_{\star} R$ serves as a precaution to keep the disk aspect ratio $h/R \gtrsim 10^{-3}$. 

\subsection{Units and dimensionless parameters} \label{sec:units}

The above set of equations can be transposed to arbitrary scales. We measured masses relative to the central mass with $G=m_{\star}=1$. We took the disk inner radius $r\lin=1$ as length unit and the inner Keplerian frequency $\Omega_{\star}\left(r\lin\right)=1$ as frequency unit. The inner Keplerian period is denoted $t\lin \equiv 2\uppi / \Omega\lin$.

A global disk model admits two parameters: its total mass\footnote{The disk mass is only used as a scaling factor. With the imposed profile $\Sigma \propto R^{-2}$, the disk mass would not converge to a finite value when extending the computational domain radially to infinity.} relative to the central mass $M\equiv m_{\mathrm{disk}} / m_{\star}$ and its dimensionless cooling time $\beta$. Additionally, the disk thickness can be characterized by its aspect ratio $h/R$, where $h$ denotes the (local) hydrostatic pressure scale height. In the following, the aspect ratio $h/R$ is always measured as a density-weighted average of $c_{s,\mathrm{iso}}/v_{\varphi}$, where $c_{s,\mathrm{iso}} \equiv \sqrt{P/\rho}$ is the isothermal sound speed --- the adiabatic sound speed is simply noted $c_s\equiv \sqrt{\gamma}c_{s,\mathrm{iso}}$. From the gas surface density $\Sigma \equiv \int_{-\infty}^{+\infty} \rho \dd z$ and epicyclic frequency $\kappa$ one can evaluate the stability of a rotationally supported disk to self-gravitating disturbances via the Toomre number:
\begin{equation} \label{eqn:toomre}
  Q\equiv \frac{\kappa c_{s,\mathrm{iso}}}{\uppi G \Sigma}.
\end{equation}
Typically, when $Q \lesssim 1$ the GI extracts orbital energy and converts it into heat via turbulence and shocks. We therefore expect gravito-turbulence to regulate itself and keep the disk close to marginal stability, that is, $Q \sim 1$.

\subsection{Theoretical background} \label{sec:theory}

The Toomre number \eqref{eqn:toomre} can decide on local linear stability: when $Q<1$ axisymmetric density disturbances grow exponentially in razor-thin disk models. For slightly larger $Q$ there exist no unstable normal modes; nonetheless the disk may be nonlinearly unstable to sufficiently large perturbations \citep{riols17}. On the other hand, there also exist global linear instabilities able to amplify eccentric \citep{adams89} or spiral structures \citep{noh91,laughlin96}. These global modes are standing waves in radius and typically rely on radial features such as boundaries or vortensity extrema \citep{papaloizou91}. In between the strictly local and global approaches is the study of tightly wound (WKBJ) density waves. WKBJ wave modes are useful to interpret the morphology and transport properties of the large-scale structures emerging in global models of GI turbulence. In this section, we adopt this intermediate approach and run through the ideas most pertinent to our simulations.

\subsubsection{Dispersion relations} \label{sec:disprel}

In the WKBJ approximation, the spiral modes of the disk may be described locally by radial and azimuthal wavenumbers $\left(k,m\right)$ such that
\begin{equation} \label{eqn:spiralfront}
  \frac{\dd R}{\dd \varphi} = -\frac{m}{k} = - R \tan\left(i\right)
\end{equation}
describes the morphology of a wave front, with $i$ the pitch angle between the spiral front and a circle around the central object. In this local approximation, linear waves in razor-thin 2D disks obey the dispersion relation
\begin{equation} \label{eqn:disp2d}
  \begin{split}
m^2 \left(\Omega_s - \Omega \right)^2 &= \kappa^2 -  2\uppi G \Sigma \vrac{k}+ c_s^2 k^2 \\
&= \kappa^2 \left(1 - Q^{-2}\right) + c_s^2 \left(k - k_0\right)^2,
  \end{split}
\end{equation}
where $\Omega\equiv v_{\varphi}/R$ is the gas angular velocity, $\Omega_s \equiv \omega/m$ is the angular pattern speed of a mode with (constant) frequency $\omega$, and $k_0 \equiv \uppi G \Sigma / c_s^2$ is the wavenumber most strongly influenced by self-gravity \citep[see, for example,][]{binney08}. When $Q<1$, there exist unstable modes whose growth rate is maximal for $\vrac{k} = k_0$. When $Q=1$, the modes with $\vrac{k}=k_0$ are marginally stable and their pattern is corotating with the gas ($\Omega_s = \Omega$). The linear theory makes no prediction for the preferred azimuthal wavenumber $m$, however.

It is customary to adopt a quasi-linear theory of GI saturation and attribute the maintenance of a marginal ($Q\approx 1$) turbulent state to the action of the marginally stable modes. Their properties are thus of considerable importance and care should be taken so that they are not mischaracterized. A critical feature of the marginal modes at $Q=1$ is that their radial wavelength equals that of the disk scale height ($k_0=1/h$), and is thus relatively short. This property brings into question the validity of the 2D disk model itself, which requires $k \ll 1/h$. It is possible to amend the dispersion relation \eqref{eqn:disp2d} so as to account for the ``dilution'' of the disk's gravity by its finite thickness:
\begin{equation} \label{eqn:disp3d}
  m^2 \left(\Omega_s - \Omega\right)^2 \simeq \kappa^2 - \frac{2\uppi G \Sigma \vrac{k}}{1 + \vrac{k} / k_0} + c_s^2 k^2
\end{equation}
\citep[see][equation 15.18]{bertin14}. The stability threshold is then shifted to $Q \lesssim 0.647$ but marginally stable modes are still corotating with the gas.

However, the disk's gravity is only one part of the picture. The ``quasi-3D'' relation \eqref{eqn:disp3d} omits the modes' vertical structure and velocities, which become increasingly important as $k$ approaches $1/h$. In nonisothermal disks especially, 3D effects can radically alter the nature of the density waves and their dispersion relations \citep{lubow98}. Moreover, when including self-gravity, \citet{mamarice10} showed that instability first appears on the (usually incompressible) inertial wave branch, a complication that may impact on its radial propagation.

Last but not least, spiral waves excited to significant amplitudes (and consequently exhibiting steep gradients) are nonlinear and thereby couple a range of $k$ and $m$ \citep{michael12}. If a nonlinear dispersion relation can be found, it may relate $\omega$ and the preponderant $\left(k,m\right)$ differently to that predicted by linear theory \citep[e.g.,][]{shu85,fromang07}. If the waves shock and thus lose energy, the linear and nonlinear relations will deviate further again. With these problems in mind, one must be wary when constructing a saturation theory, and in particular simulation diagnostics, that rely too heavily on the details of the dispersion relation \eqref{eqn:disp3d}. 

\subsubsection{Local turbulence and large-scale spirals}

One contentious aspect of gravito-turbulence is whether it can be treated as a local process, coupling the local gas accretion and heating rates as in an $\alpha$ disk \citep{balbuspap99}. Spiral density waves that carry energy over large distances before damping are central to this problem. From the 2D linear theory of WKBJ modes, the ratio of the wave to turbulent energy fluxes is simply $\xi = \vrac{(\Omega_s-\Omega)/\Omega}$ \citep{cossins09}. Hence the importance of nonlocal energy transport associated with a given wave depends on its proximity to its corotation radius --- at which it is marginally stable. But in a quasi-linear theory of GI saturation, gravito-turbulence at any given radius is controlled by the local marginally stable modes. If these modes travel radially they will leave the region that can sustain them and then scatter and dissipate on more vigorous modes at their new location. One might then expect that WKBJ waves never travel far from their corotation radii and thus never contribute appreciably to nonlocal energy transport \citep{gammie01}. 

We are then left with a picture of gravito-turbulence that comprises an incoherent patchwork of modes with short radial wavelengths and localized to their corotation radii. This picture contrasts with the large-scale spirals reported by numerical simulations and often taken as a starting point to test the GI hypothesis in observed circumstellar disks \citep[e.g.,][]{forgan18}. One way to resolve this tension is to posit that at any given radius there exist several spiral wave packets distributed across the $2\uppi$ in azimuth, each corotating with the background orbital flow and each exhibiting the same $m$ and $k = k_0 \approx 1/h$. Over the course of several orbital periods, localized spirals on neighboring radii will intermittently coalesce and form loose composite structures on a larger radial scale. The apparent coherence of these larger structures will issue from the similar pitch angles of their constituent wakes. If there are a sufficient number of localized wave packets per radius, any snapshot of a simulation will likely show up a sequence of transient agglomerations over several radii of this type. Later in this paper we test this idea.

\subsection{Numerical method} \label{sec:numethod}

We used the version 4.3 of the \textsc{Pluto} code \citep{mignone07} to integrate Eqs. \eqref{eqn:dtrho}-\eqref{eqn:dtP} in time. We calculated solutions of the self-gravitating disk dynamics for the initial and boundary conditions described below. 

\subsubsection{Computational domain} \label{sec:domain}

The computational domain was defined by intervals in spherical coordinates. The radial interval $r \in \left[r\lin,32 r\lin\right]$ was meshed with 518 grid cells with a uniform $\log\left(r\right)$ sampling. The interval in colatitude $\theta \in \left[\uppi/2\pm 0.35 \right]$ was meshed with two different grid spacings. We uniformly meshed the midplane region $\theta \in \left[ \uppi/2\pm0.05\right]$ with $32$ cells. We meshed the extreme $\theta$ intervals with $32$ ``stretched'' cells on both sides of the midplane, whose angular increment $\Delta \theta$ increases geometrically from $\theta=\uppi/2\pm 0.05$ to the boundaries $\theta = \uppi/2\pm 0.35$. The azimuthal interval $\varphi \in \left[0,2\uppi\right]$ was uniformly meshed with 512 grid cells, corresponding to the highest value considered in the resolution studies of \citet{michael12} and \citet{steiman13}. 

\subsubsection{Integration scheme} \label{sec:scheme}

We used \textsc{Pluto} to integrate Eqs. \eqref{eqn:dtrho}-\eqref{eqn:dtP} in conservative form via a finite-volume spatial discretization and a second-order Runge-Kutta time stepping. We used a piecewise linear reconstruction \citep{mignone14} with Van Leer's slope limiter \citep{vanleer74} to determine the state of the primitive variables $\left(\rho,\bm{v},P\right)$ on both sides of each cell interface. These states are converted into conservative variables $\left(\rho,\rho \bm{v},E\right)$, where $E \equiv \rho \left(\epsilon + v^2/2\right)$ is the gas energy density. The energy equation as integrated by \textsc{Pluto} can be written
\begin{equation} \label{eqn:dtE}
  \partial_t E = -\nabla\cdot \left[\left(E + P\right)\bm{v} \right] - \nabla\cdot \left[\Phi\rho \bm{v} \right] - \Phi \partial_t \rho - q^-,
\end{equation}
rightfully excluding the time derivative $\partial_t\Phi$ of the gravitational potential. We used the HLLc Riemann solver of \citet{toro94} to compute the interface fluxes from which the volume-averaged conservative variables are updated. The gas pressure is finally recovered as $P/\left(\gamma-1\right) = E - \rho v^2/2$, independent of the gravitational potential energy. The Rankine-Hugoniot jump conditions are built into the Riemann solver, so shock heating is handled without explicit dissipation terms in Eq. \eqref{eqn:dtE}. This integration scheme is stable with the chosen Courant-Friedrichs-Lewy coefficient of $0.3$. 

Because solving Poisson's equation \eqref{eqn:poisson} can be computationally costly, we updated the gravitational potential of the gas $\Phi_{\mathrm{disk}}$ only once per time step --- not at every Runge-Kutta substep. While the integration scheme becomes formally first-order accurate in time, it is still able to properly resolve the growth of the GI. We present in Appendix \ref{app:poisson} how Poisson's equation was numerically solved and the related tests. 

\subsubsection{Initial conditions} \label{sec:initcond}

We wish to resolve the vertical structure of the disk and to make the dimensionless properties of turbulence independent of radius. The first point above requires that the disk thickness $h$ should be resolved by several grid cells at every radius, that is, $c_{s,\mathrm{iso}} / v_{\varphi} \gtrsim 10^{-2}$ near the disk midplane. The second point requires the dimensionless parameters to be independent of radius, including the disk aspect ratio $h/R$ and the Toomre number $Q$. The angular frequency of a Keplerian disk is $\Omega_{\star} \propto R^{-3/2}$, so the aspect ratio is constant when $c_{s,\mathrm{iso}} \propto R^{-1/2}$. Assuming that gravito-turbulence will maintain $Q\approx 1$ in quasi-steady state, we started near $Q = 1$ at every radius to avoid a long initial cooling phase and to keep the disk geometrically thick. This choice implies $\Sigma \propto R^{-2}$ for the initial gas surface density. 

We set the initial $\Sigma(R) = \Sigma\lin \left(R/r\lin\right)^{-2}$ such that $m_{\mathrm{disk}}\left(x\right) \equiv \int_{r\lin}^{x} 2\uppi \Sigma R \dd R$ equals the prescribed disk mass at the outer radius $x=r\lout$. The initial velocity was set to $v_r=v_{\theta}=0$ and
\begin{equation}
  v_{\varphi}\left(R\right) = \sqrt{\frac{G \left[m_{\star} + m_{\mathrm{disk}}\left(R\right)\right]}{R}}.
\end{equation}
The isothermal sound speed $c_{s,\mathrm{iso}}$ was initialized such that $\Omega c_{s,\mathrm{iso}} / \uppi G \Sigma = 1$ at every radius. We then distributed the mass density as a simple Gaussian in the vertical direction:
\begin{equation}\label{eqn:rho0}
  \rho_0\left(R,z\right) \equiv \frac{\Sigma\left(R\right)}{h\left(R\right)\sqrt{2\uppi}} \exp\left[-\frac{1}{2}\left(\frac{z}{h\left(R\right)}\right)^2\right].
\end{equation}
These initial conditions are not exactly in quasi-static equilibrium due to omitting the radial pressure gradient and the vertical component of the gas self-gravity in the momentum equation. As we show below, the disk quickly relaxes to a quasi-steady state with no violent redistribution of mass or momentum.

\subsubsection{Boundary conditions} \label{sec:boundcond}

Let $f\left(x_{\mathrm{bound}}-x\right) = \pm f\left(x_{\mathrm{bound}}+x\right)$ represent symmetric ($+$) or antisymmetric ($-$) conditions for the flow variable $f$ near the boundary of coordinate $x_{\mathrm{bound}}$. We used the same boundary conditions at the inner and outer radial boundaries: we symmetrized $\rho$ and $v_{\varphi}/r$, while we antisymmetrized $r^2 v_r$ and $v_{\theta}$. These conditions favor solid-body rotation and disfavor radial in and outflows. On the colatitude $\theta$ boundaries we symmetrized $\rho$ and antisymmetrized all three velocity components. These conditions also prevent flows through the domain boundaries, and they disfavor rotational support. Finally, we imposed a small but nonzero pressure in the ghost cells of every boundary, corresponding to the pressure floor given in the section below. This choice prevents internal energy inputs from the boundaries.

These boundary conditions are meant to be simple and yet allow the disk to find a quasi-steady state inside the computational domain. We note that the (anti-)symmetry cannot be imposed exactly due to the nonuniform grid spacings and to the reconstruction and conversion steps involved in the finite-volume scheme. As a consequence, the total mass inside the domain does evolve by a few per cent per hundred inner orbits (see Appendix \ref{app:inertia}), not affecting our conclusions. 

\subsubsection{Additional precautions} \label{sec:precautions}

The vertical density stratification and the presence of sonic turbulence are challenging to resolve with a stable and accurate scheme. We took additional precautions to maintain numerical stability while retaining second-order spatial accuracy over most of the computational domain. We imposed a density floor $\rho_{\mathrm{floor}} = 10^{-6} \rho_0\left(r_{\mathrm{out}},0\right)$ from the initial conditions \eqref{eqn:rho0}. Since the initial midplane density decreases as $R^{-3}$, this floor only intermittently affects a few grid cells at the edges of the computational domain. We imposed a pressure floor $P_{\mathrm{floor}} = \rho_{\mathrm{floor}} c_{\mathrm{floor}}^2\left(r_{\mathrm{out}}\right)$, where $c_{\mathrm{floor}}\left(r_{\mathrm{out}}\right)$ is the threshold sound speed used in the cooling function \eqref{eqn:cool}. This pressure floor also affects a small fraction of grid cells located near the radial boundaries of the domain and in the vicinity of strong discontinuities. The density and pressure floors being applied independently, the gas temperature may artificially change in the concerned grid cells. As a precaution, we prevented the isothermal sound speed from going below $c_{\mathrm{floor}}\left(R\right)$. These floor values are chosen to have a negligible effect on the disk mass and internal energy. Finally, should the relative pressure difference between neighboring cells exceed a factor five, we switched to the more dissipative MINMOD slope limiter \citep{roe86} and to a two-state HLL Riemann solver \citep{harten83} at the concerned cell interfaces. This switch affects between one and six per cent of all grid cells, decreasing with increasing disk thickness. Being restricted to the vicinity of shocks, it does not enhance artificial heating in smooth regions of the flow. To minimize the influence of such procedures we favor thick and vertically well-resolved disks. 

\subsection{Data reduction} \label{sec:reduction}

We used various procedures to analyze the simulation data. We only present the main ones here, leaving more elaborate measurement techniques to be introduced in later sections when needed.

Let $X$ be a flow quantity varying in time and over the simulation grid. The midplane distribution of $X$ is denoted as $X_{\mathrm{mid}}\left(r,\varphi\right)$. The time average of $X$ is denoted as $\overline{X}$ and taken over one outer Keplerian period $2\uppi / \Omega_{\star}\left(r\lout\right) \approx 181 t\lin$. Spatial averages are denoted with brackets and defined as a volume-weighted sum over grid cells in the corresponding dimensions. For example, affecting the indices $\left(i,j,k\right)$ to cell attributes at coordinates $\left(r_i,\theta_j,\varphi_k \right)$, the azimuthal average of $X$ is:
\begin{equation}
  \brac{X}_{\varphi}\left(r_i,\theta_j\right) = \frac{\sum_{\varphi_k=0}^{\varphi_k<2\uppi} \left[ X \,\delta V\right]_{i,j,k}}{\sum_{\varphi_k=0}^{\varphi_k<2\uppi} \,\delta V_{i,j,k}},
\end{equation}
where $\delta V$ denotes the volume of a grid cell. Looking at geometrically thin disks on a spherical grid, we opted to use $\brac{X}_{\theta}$ in place of true vertical averages for simplicity. Radial averages are taken over $r/r\lin \in \left[2,16\right]$ and evaluated with a $\dd\!\log\left(r\right)$ measure matching the logarithmic grid spacing. We define a density-weighted vertical average as 
\begin{equation}
  \brac{X}_{\rho}\left(r_i,\varphi_k\right) = \frac{\sum_{\theta_j} \left[ X \rho \,\delta V\right]_{i,j,k}}{\sum_{\theta_j} \left[\rho \,\delta V\right]_{i,j,k}}. 
\end{equation}
From the vertically and azimuthally averaged profile $\brac{X}_{\theta,\varphi}$, we define the relative fluctuations
\begin{equation}
  \tilde{X}\left(r,\varphi\right) = \frac{ \brac{X\left(r,\theta,\varphi\right) - \brac{X}_{\theta,\varphi}\left(r\right)}_{\theta}}{\brac{X}_{\theta,\varphi}\left(r\right) }.
\end{equation}

\section{Reference simulation} \label{sec:refsim}

We start by examining one simulation in detail, with a disk mass $M=1/3$ of the central mass and a local cooling time $\beta=10$. We label this simulation M3B10 and take it as our reference case for later comparison. We sample different disk masses and cooling times in Sect. \ref{sec:sample}.

\subsection{Initial transient and disk spreading}

\begin{figure}
  \centering
  \includegraphics[width=\columnwidth]{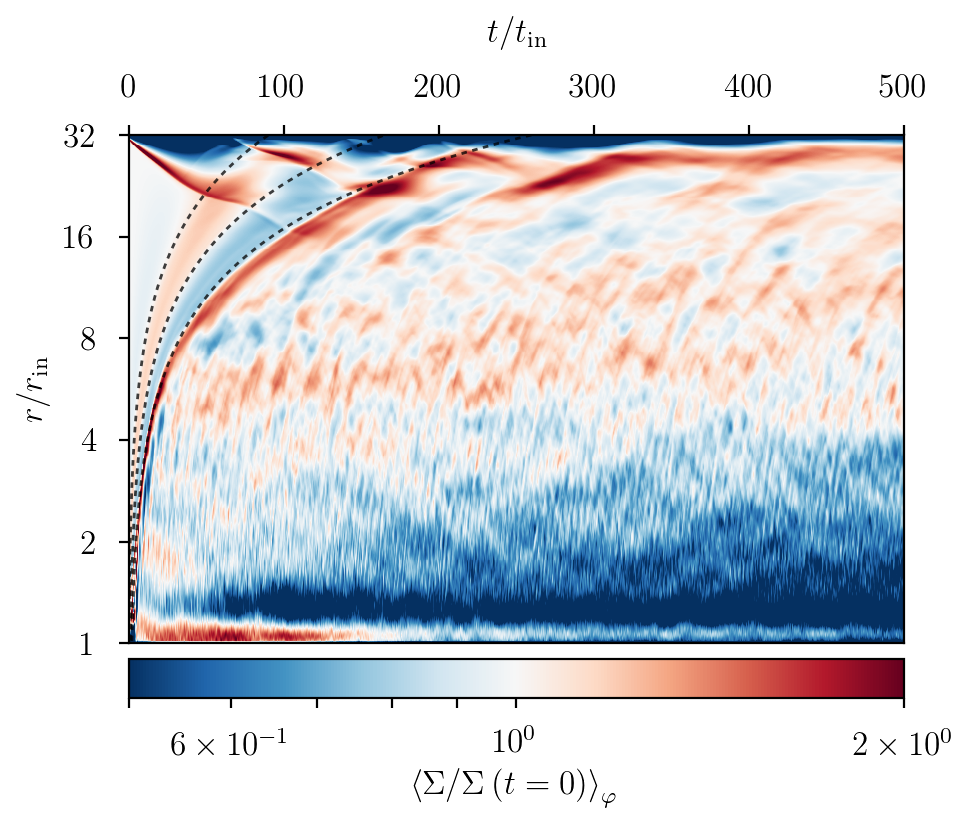}
  \caption{Evolution in time (horizontal axis) of the radial profile (vertical axis) of the gas surface density relative to its initial profile in run M3B10. The dashed curves on the left fit the initial transient fronts with $r\left(t \right) \propto t^{2/3}$, i.e., a radial velocity proportional to $r^{-1/2}$.} 
  \label{fig:m3b10_map_dn}
\end{figure}

We show in Fig. \ref{fig:m3b10_map_dn} the evolution in time of the azimuthally averaged surface density $\brac{\Sigma}_{\varphi}$ relative to the initial profile $\Sigma\left(R,t=0\right)$. Because the initial conditions are not in quasi-static equilibrium, the disk first contracts vertically to restore momentum balance. From a local perspective, the time needed to restore a hydrostatic equilibrium is $\sim h/c_s \simeq \Omega_{\star}^{-1}$, shorter than the imposed cooling time $\beta \Omega_{\star}^{-1}$. The radius at which the product $\Omega_{\star} t = 1$ increases as $t^{2/3}$, so the contraction front appears to propagate outward at a velocity $\propto r^{-1/2}$. From a global perspective, this differential contraction sweeps material radially and induces an inertial-acoustic response visible in the gas surface density, as traced by the dashed curves in Fig. \ref{fig:m3b10_map_dn}. 

After the initial transient, we can distinguish narrow stripes of axisymmetric density perturbations traveling radially at roughly Keplerian velocities. These fluctuations evolve on local orbital timescales while the global redistribution of mass throughout the disk takes several $10^2 t\lin$. The disk has therefore reached a quasi-steady state thermally, at least in the inner $r/r\lin\lesssim 16$, and it is the dissipation of turbulent motions that balances cooling losses. During this phase, the surface density decreases at small radii and increases at large radii, varying by less than a factor two relative to the initial profile on most of the radial interval. At $500 t\lin$ the disk has lost fourteen per cent of its initial mass, equivalent to the mass initially located inside $1.6 r\lin$. Most of this mass has left the computational domain through the inner radial boundary, while the density increase in $r/r\lin \in \left[4, 23\right]$ can only result from an outward transport of mass. We explain both trends in Sect. \ref{sec:angmflx} by measuring the radial flux of angular momentum.

\subsection{Vertical structure of the disk} \label{sec:m3vert}

\begin{figure}
  \centering
  \includegraphics[width=\columnwidth]{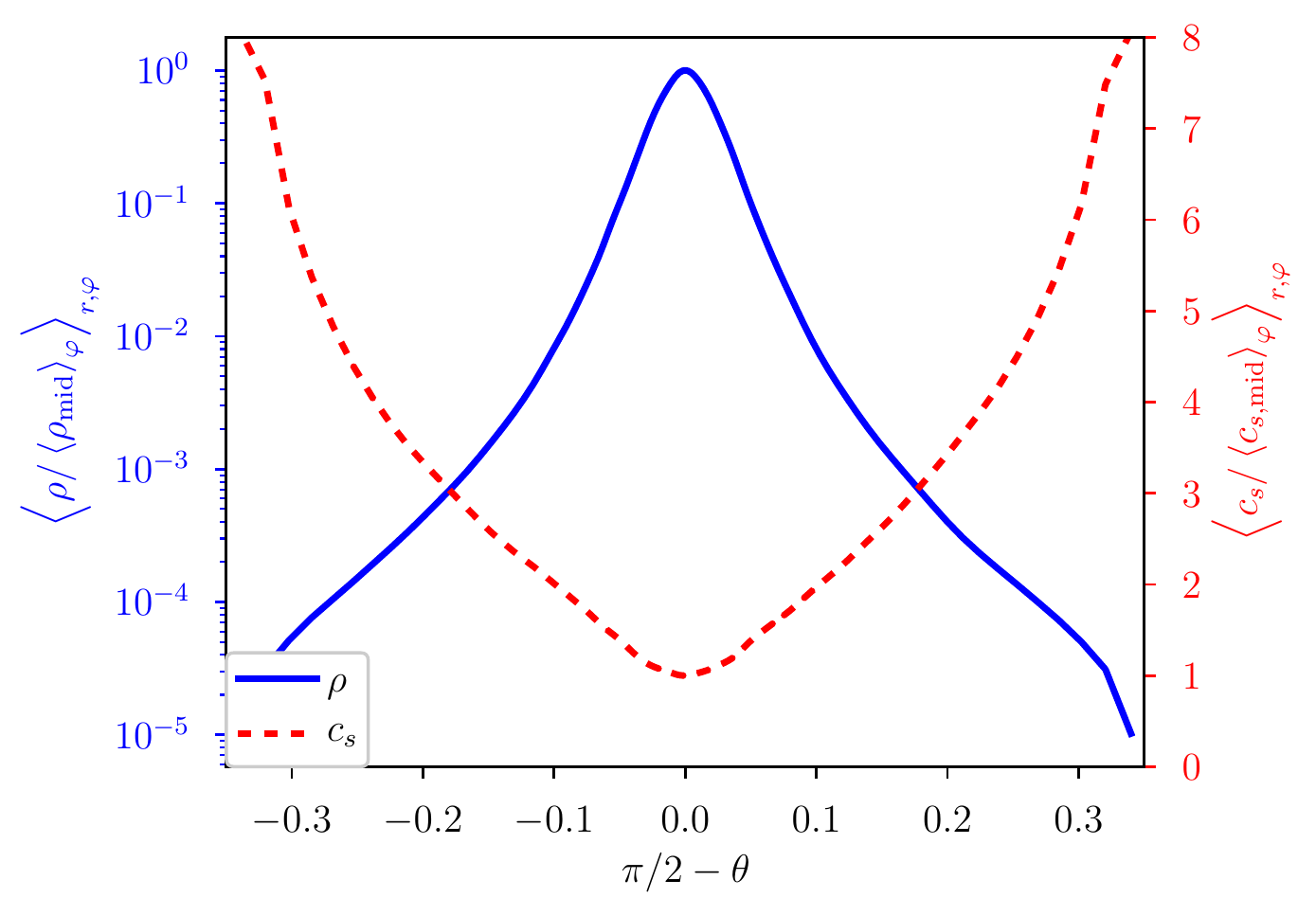}
  \caption{Radial and azimuthal averages of the density (blue, left axis) and sound speed (red, right axis) relative to their azimuthally averaged midplane value at time $500 t\lin$ in run M3B10.}
  \label{fig:m3b10_COl_rho}
\end{figure}

In the midplane of run M3B10 we measure $c_{s,\mathrm{iso}}/v_{\varphi} \approx 0.04$ upon radial and azimuthal averaging, so we resolve one midplane pressure scale height by roughly thirteen grid cells. If the disk was vertically isothermal, the computational domain would cover eight pressure scale heights on both sides of the midplane. As a result, we are well placed to probe the vertical structure and time-dependent dynamics of the disk. 

For every grid cell at time $500 t\lin$, we divide the local density and sound speed by their azimuthally averaged midplane value at the same spherical radius. We then average these ratios over $\varphi \in \left[0,2\uppi\right]$ and $r/r\lin \in \left[2 ,16 \right]$. The resulting profiles represent the mean vertical structure of the disk and are drawn in Fig. \ref{fig:m3b10_COl_rho}. 

The gas density decreases by roughly five orders of magnitude with height at any radius. The density profile is not Gaussian --- as initially prescribed --- but is instead narrower in the midplane with flat tails at high latitudes. In the midplane region $\vrac{\uppi/2 - \theta} \lesssim 0.05$, the gas density is primarily compressed by the disk's own gravity. Indeed, the vertical component of the gravitational acceleration is approximately five times larger than the central $\Phi_{\star}\left(r\right)$ contribution at small heights.

Away from the midplane, the sound speed increases with height by up to a factor eight, allowing vertical hydrostatic balance with a density profile flatter than Gaussian. This temperature increase is much steeper than what \citet[][figure 1]{shi14} and \citet[][figure 2]{riols17} reported from shearing box experiments with the same cooling prescription. We believe that these differences issue from our imposed boundary conditions. By antisymmetrizing $v_{\varphi}$ at the $\theta$ boundaries we hinder rotational support at high latitudes, and to reach a steady state the gas can either travel radially or become pressure supported. By antisymmetrizing $r^2 v_r$ at the radial boundaries we prevent the development of Bondi-Parker flows, leaving pressure to support the gas against gravity in the radial direction. The combination of a decreasing density and increasing temperature implies that the entropy quickly increases with height, stabilizing the disk against convective motions \citep{boley06b}.

After azimuthal averaging, we measure radial infall velocities at high latitudes. We find $\brac{v_r}_{\varphi} \approx -0.13 \brac{c_s}_{\varphi}$ at $\vrac{\uppi/2-\theta} = 0.25$ and reaching $\brac{v_r}_{\varphi} \approx -0.35 \brac{c_s}_{\varphi}$ near the $\theta$ boundaries. This mean inflow remains much slower than free fall, confirming that pressure makes up for the lack of rotational support in the disk corona. The turbulent motions being subsonic above the disk, the heat source is presumably localized near the midplane and heat is transported by the vertical velocity fluctuations $\gtrsim 0.1 c_s$ to the corona. Despite the temperature increase with height, the internal energy stored above $\vrac{\uppi/2 - \theta} > 0.1$ represents only two per cent of the total internal energy of the disk. Since the corona extracts a negligible amount of heat from the background rotation, we do not expect it to significantly influence the saturation level of the GI. 

\subsection{Instantaneous equatorial flow} \label{sec:insteqf}

\begin{figure}
  \centering
  \includegraphics[width=\columnwidth]{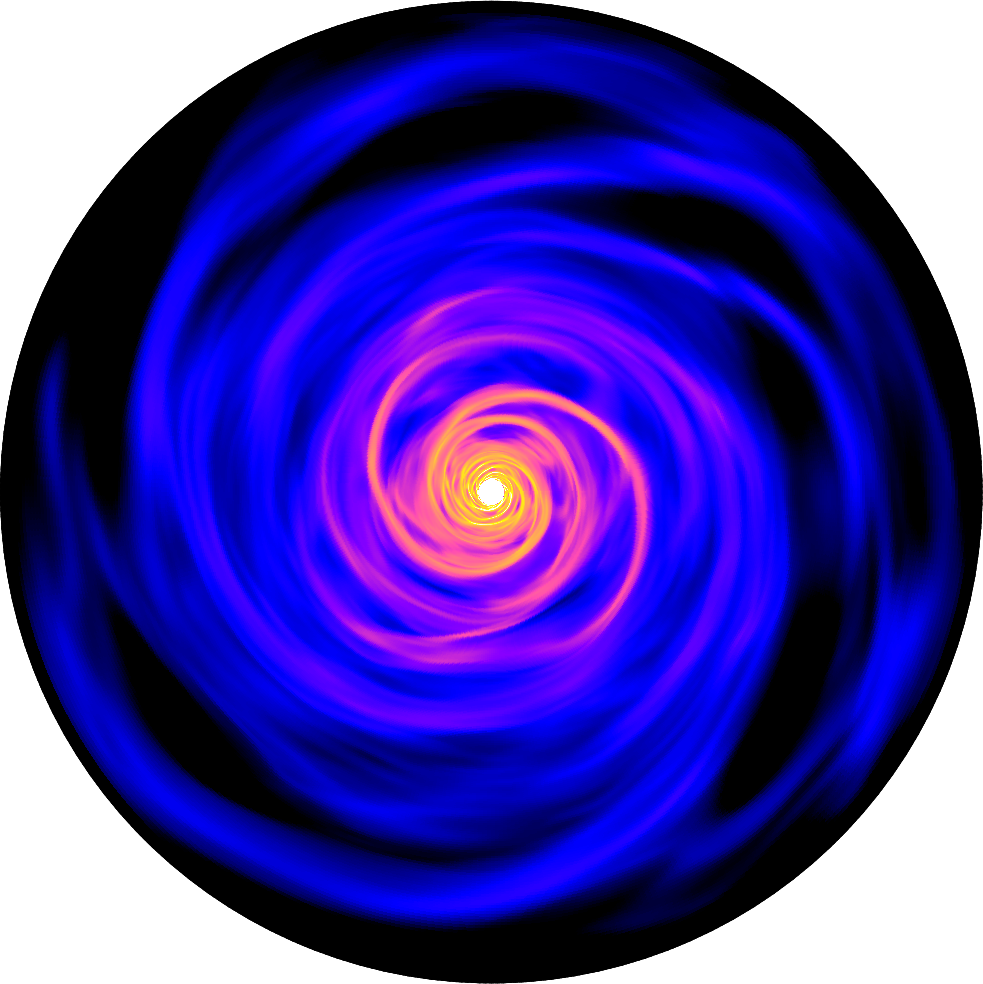}
  \caption{Gas surface density $\Sigma$ at time $500 t\lin$ in run M3B10; the color scale spans three orders of magnitude from black to white.}
  \label{fig:m3b10_EQt_sig}
\end{figure}

\begin{figure}
  \centering
  \includegraphics[width=\columnwidth]{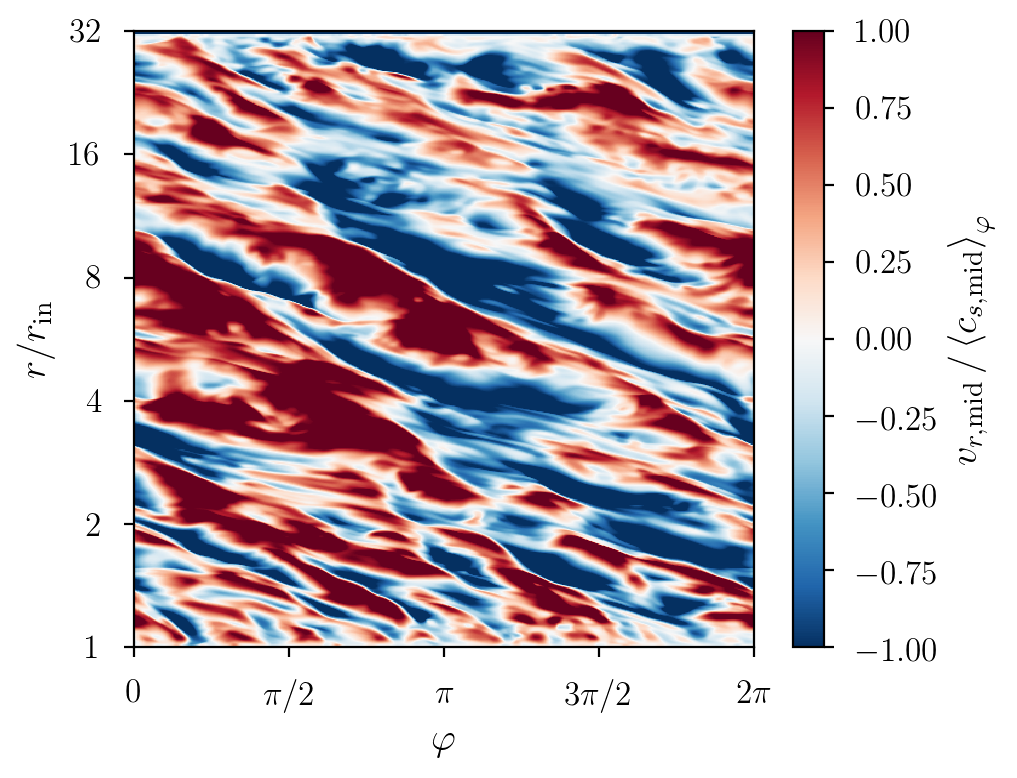}
  \caption{Radial velocity relative to the azimuthally averaged sound speed in the midplane of run M3B10 at time $500 t\lin$. The velocity fluctuations reach supersonic amplitudes both outward (red) and inward (blue).}
  \label{fig:m3b10_EQm_vrcs}
\end{figure}

We now look at the instantaneous equatorial flow in run M3B10 at time $500 t\lin$, starting traditionally with the gas surface density in Fig. \ref{fig:m3b10_EQt_sig}. The disk is clearly nonaxisymmetric, which Fig. \ref{fig:m3b10_map_dn} could not capture because of the azimuthal averaging. The gas density is organized in spiral arms with a relatively tight winding and order-unity density fluctuations along $\varphi$ at any radius. The largest spiral arms span more than $\uppi$ in azimuth, and some spirals extend to several times their footpoint in radius. However, we are not able to identify a clear-cut global spiral pattern by visually inspecting Fig. \ref{fig:m3b10_EQt_sig}.

To examine these structures quantitatively, we move on to the $\left(r,\varphi\right)$ plane and show the midplane distribution of the radial velocity in Fig. \ref{fig:m3b10_EQm_vrcs}. This figure reveals transsonic motions at all radii. The velocity fluctuations are oriented both inward and outward in roughly equal proportions. We identify spiral wakes as the oblique lines where the velocity $v_r=0$ changes sign in a compressive fashion. Bearing in mind the logarithmic vertical scale of Fig. \ref{fig:m3b10_EQm_vrcs}, these wakes appear to have a roughly logarithmic spiral morphology. Across each wake, the velocity jumps by more than twice the mean sound speed. We deduce that the flow must shock in the vicinity of the wakes, making them a primary location of entropy generation. 

Because Fig. \ref{fig:m3b10_EQm_vrcs} only represents the instantaneous state of the flow, it is too early to describe the spiral structures as global waves propagating coherently through the disk. To address this issue, we measure their pattern speed and temporal coherence in Sects. \ref{sec:eqmotion} \& \ref{sec:properwake}. Because Fig. \ref{fig:m3b10_EQm_vrcs} only represents a slice of the disk midplane, we also cannot assume that the flow shocks over the entire height of the disk. We show in Sect. \ref{sec:merryglyph} that the compressive midplane motions in fact transit to vertical outflows near the wakes.

\subsection{Toomre instability}

\begin{figure}
  \centering
  \includegraphics[width=\columnwidth]{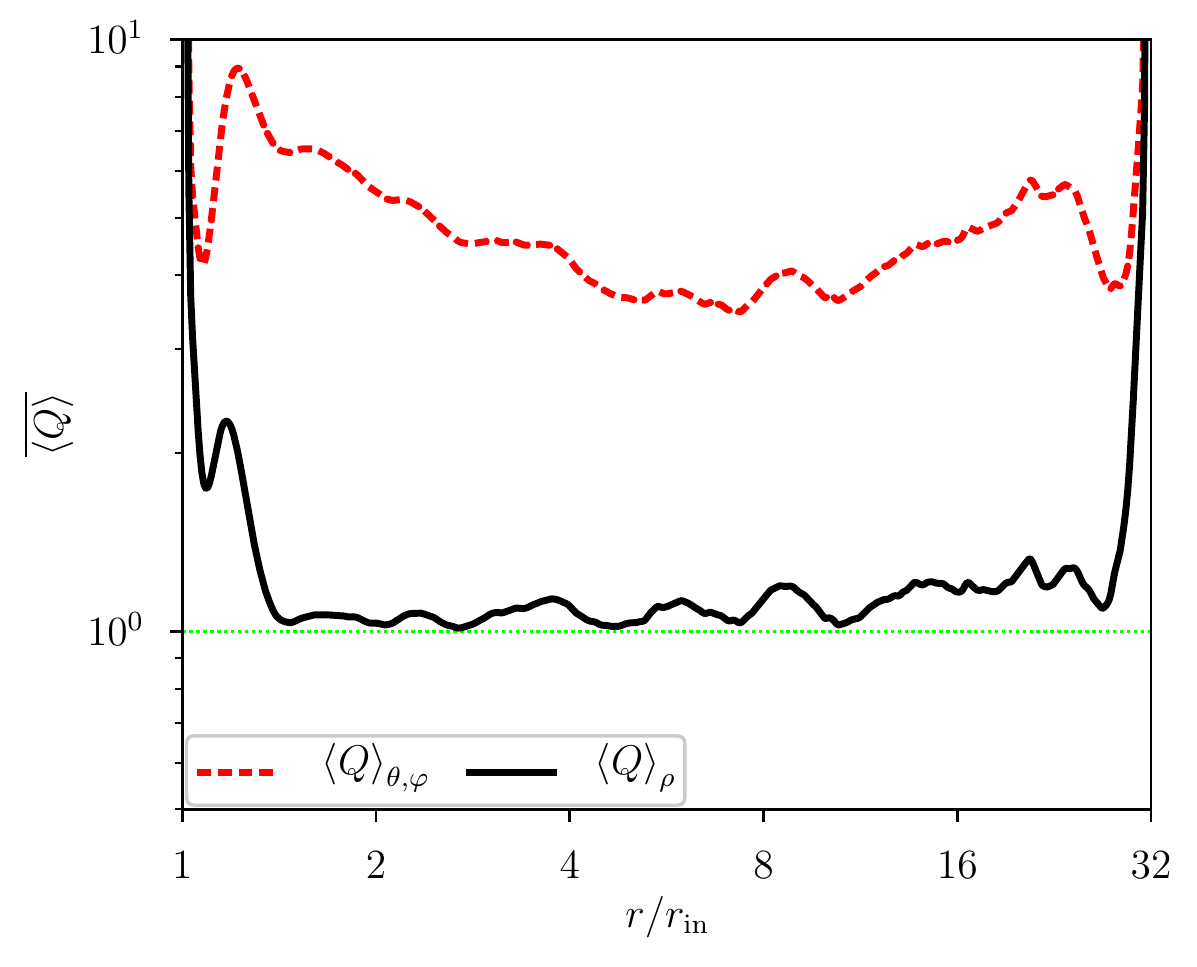}
  \caption{Radial profiles of the time-averaged Toomre number $Q$ in run M3B10, using either the density-weighted sound speed $\brac{c_{s,\mathrm{iso}}}_{\rho}$ (solid black) or a simple vertical averaging (dashed red) in Eq. \eqref{eqn:toomre}. The dotted horizontal line marks the threshold $Q=1$ of marginal stability.}
  \label{fig:m3b10_avg_tom}
\end{figure}

Figure \ref{fig:m3b10_EQm_vrcs} shows both small-scale fluctuations and large-scale spirals with transsonic amplitudes. To confirm the predominant role of the GI in driving this turbulence, we measure the Toomre number in the quasi-steady state of run M3B10. We used the density-weighted mean orbital velocity $\Omega_{\rho} \equiv \brac{v_{\varphi}/R}_{\rho}$ to estimate the epicyclic frequency
\begin{equation}
\kappa \simeq \sqrt{ \frac{2\Omega_{\rho}}{R} \frac{\dd R^2\Omega_{\rho}}{\dd R}}
\end{equation}
in the definition \eqref{eqn:toomre} of the Toomre number. To distinguish the cold and dense disk midplane from its warm corona, we compute the mean sound speed with and without density weighting, denoting the corresponding Toomre numbers as $\brac{Q}_{\rho}$ and $\brac{Q}_{\theta,\varphi}$ respectively. The profiles of $Q$ are averaged over one outer Keplerian period and drawn in Fig. \ref{fig:m3b10_avg_tom}.

With a simple vertical averaging, the Toomre number $\brac{Q}_{\theta,\varphi} \approx 5$ is significantly larger than unity. However, this value is dominated by the warm corona above the disk midplane, see Fig. \ref{fig:m3b10_COl_rho}, and is thus unsuitable. If we instead focus on the midplane by taking the density-weighted mean Toomre number, we find $\brac{Q}_{\rho} \approx 1$ for almost all radii, with fluctuations of less than thirty per cent after time-averaging \citep{rafikov15}. Since $Q=1$ corresponds to marginal gravitational stability \citep{toomre64}, the measured profile $\brac{Q}_{\rho} \approx 1$ verifies that the GI is the main driver of turbulence: if a region of the disk cools below $Q < 1$, the instability converts orbital energy into heat to restore $Q\gtrsim 1$. We therefore expect the orbital energy extraction rate, and a fortiori the angular momentum flux, to depend on the prescribed $\beta$ cooling timescale \citep{gammie01}. 

\subsection{Absence of other instabilities}

The saturation of the GI could be complicated by other processes such as hydrodynamic instabilities unrelated to the self-gravitating nature of the disk, or even by unsuitable boundary conditions \citep{shu90,linpap11}. We attempt to rule out several common possibilities by applying simple stability criteria to the quasi-steady state of run M3B10.

We can discard the vertical shear instability \citep{urpin98} given the relatively long cooling times $\beta > 1$ and the stabilizing effect of the steep vertical entropy stratification \citep{nelson13}. On the other hand, while there is a vertical shear in the mean radial infall, it is only found high above the disk and is unlikely to affect the flow near the disk midplane. 

\begin{figure}
  \centering
  \includegraphics[width=\columnwidth]{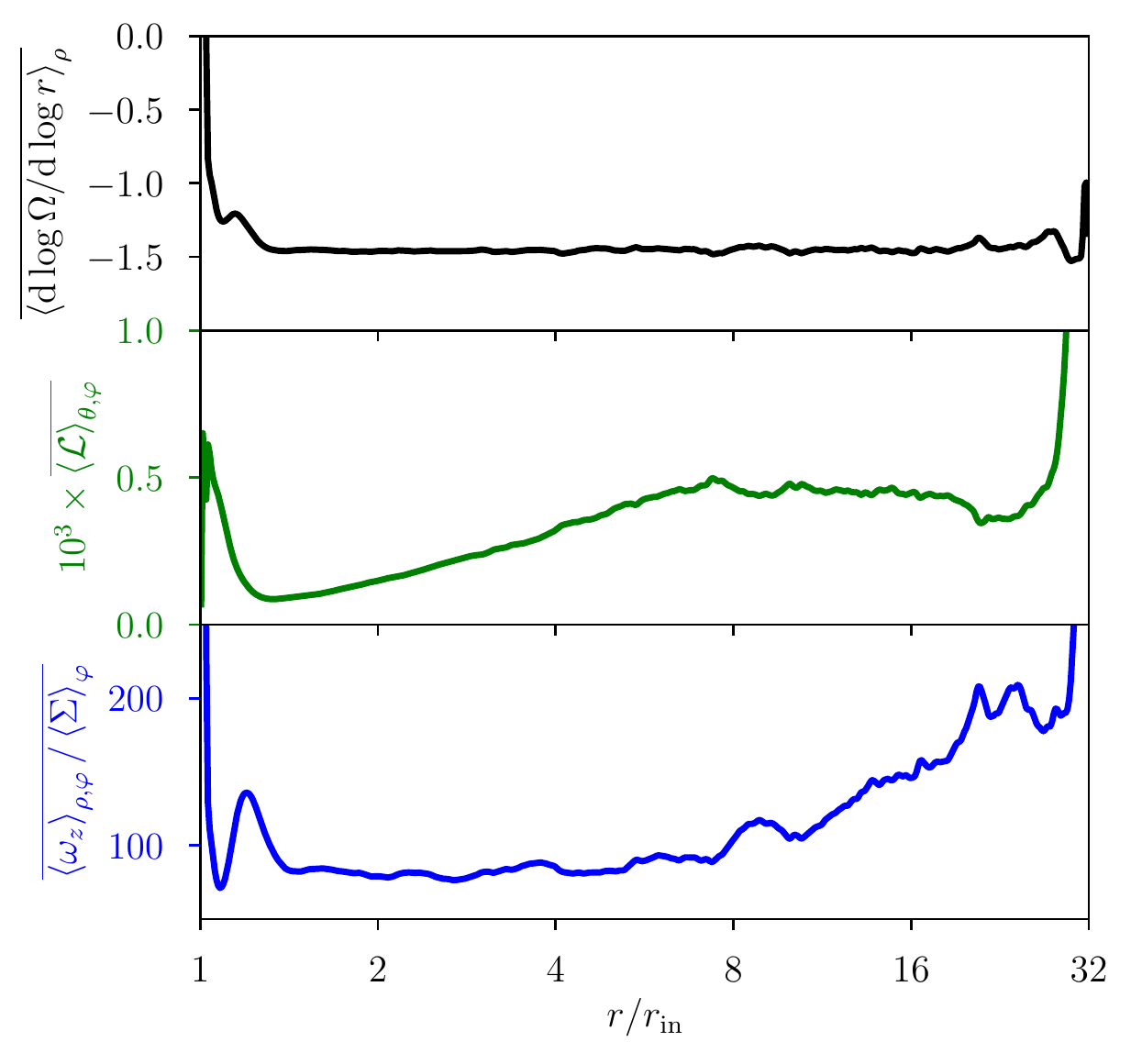}
  \caption{time-averaged radial profiles of three key quantities to assess the disk stability: the dimensionless shear-rate (upper panel), the function $\mathcal{L}$ introduced by \citet{lovelace99} and defined by \eqref{eqn:rossby}, and the vertical component of the vortensity.}
  \label{fig:m3b10_avg_stab}
\end{figure}

On the top panel of Fig. \ref{fig:m3b10_avg_stab} we draw the dimensionless shear rate $q \equiv \brac{\dd\!\log \Omega / \dd\!\log r}_{\rho}$. Apart from the inner radius where the boundary conditions enforces solid body rotation ($q=0$), the shear profile remains Keplerian througout the disk ($q=-3/2$) and never approaches Rayleigh's criterion $q<-2$, allowing us to discount the centrifugal instability.

To test for the presence of the Rossby wave instability, we followed \citet{lovelace99} and computed the function 
\begin{equation} \label{eqn:rossby}
  \mathcal{L}\left(R\right) \equiv \brac{\frac{\Sigma}{\omega_z} \left(\frac{P}{\Sigma^{\gamma}}\right)^{2/\gamma}}_{\theta,\varphi},
\end{equation}
using $\omega_z \simeq \left(1/R\right) \dd \left(R^2 \Omega_{\rho}\right)/\dd R$ to estimate the mean vertical vorticity. As shown on the middle panel of Fig. \ref{fig:m3b10_avg_stab}, this function is maximal at the radial boundaries of the domain and has no extrema in the main body of the disk, thereby dismissing the Rossby wave instability as a source of waves and turbulence, except possibly near the inner radius.  

Finally, we show the radial profile of the vortensity $\omega_z / \Sigma$ on the bottom panel of Fig. \ref{fig:m3b10_avg_stab}. Though the flow is nonbarotropic and the vortensity is not conserved, it may nonetheless suggest nonaxisymmetric instability \citep{papaloizou91}. This vortensity profile features a small bump near the inner radius and a relatively smooth increase toward the outer radius. The bump could contribute to exciting nonaxisymmetric waves inside $r/r\lin<1.4$, though its influence would likely remain localized near the inner boundary. Over the rest of the disk, the absence of a clear vortensity extremum discounts instabilities of nonaxisymmetric waves corotating with the extremum. 

We acknowledge that our application of stability criteria to a fully nonlinear turbulent state is neither rigorous nor exhaustive. It nevertheless supports the idea that our simulations exhibit a relatively pure form of global gravito-turbulence. 

\subsection{Angular momentum transport} \label{sec:angmflx}

Mass accretion is governed by the equation of angular momentum transport, which can be put in conservative form even for self-gravitating disks \citep{lbk72}. The radial flux of angular momentum through the disk consists of two terms, one hydrodynamic and the other gravitational. The hydrodynamic (Reynolds) stress arises from correlated velocity fluctuations and can be defined as
\begin{equation} \label{eqn:Rrp}
  \mathrm{R}_{R\varphi} = \rho \left( v_R - \brac{v_R}_{\rho} \right) \left( v_{\varphi} - \brac{v_{\varphi}}_{\rho} \right).
\end{equation}
The gravitational stress comes from correlations between the $R$ and $\varphi$ gravitational accelerations and takes the form
\begin{equation} \label{eqn:Grp}
  \mathrm{G}_{R\varphi} = \frac{\left(\partial_R \Phi_{\mathrm{disk}}\right) \left(\partial_{\varphi}\Phi_{\mathrm{disk}}\right)}{4\uppi R G}.
\end{equation}
We normalize the stress tensors by the vertically averaged gas pressure to obtain dimensionless viscosity coefficients:
\begin{equation} \label{eqn:alpha}
  \alpha_{\mathrm{R}} \equiv \frac{\mathrm{R}_{R\varphi}}{\brac{P}_{\theta,\varphi}}; \qquad \alpha_{\mathrm{G}} \equiv \frac{\mathrm{G}_{R\varphi}}{\brac{P}_{\theta,\varphi}}; \qquad \alpha \equiv \alpha_{\mathrm{R}} + \alpha_{\mathrm{G}}.
\end{equation}
Assuming that the orbital energy extracted by the stress is locally dissipated into heat, the condition of thermal balance against $\beta$ cooling
(here denoted by LTE) uniquely determines the value of $\alpha$ \citep{gammie01}. With the chosen normalization\footnote{The different choice of normalization used by \citet{gammie01} introduces an additional $\gamma \left(\dd\!\log \Omega / \dd\!\log R\right)$ in the denominator.}, this value is:
\begin{equation} \label{eqn:alpha0}
  \alpha_{\mathrm{LTE}} = \frac{1}{\left(-\frac{\dd\!\log \Omega}{\dd\!\log R}\right) \left(\gamma - 1\right) \beta}. 
\end{equation}

\begin{figure}
  \centering
  \includegraphics[width=\columnwidth]{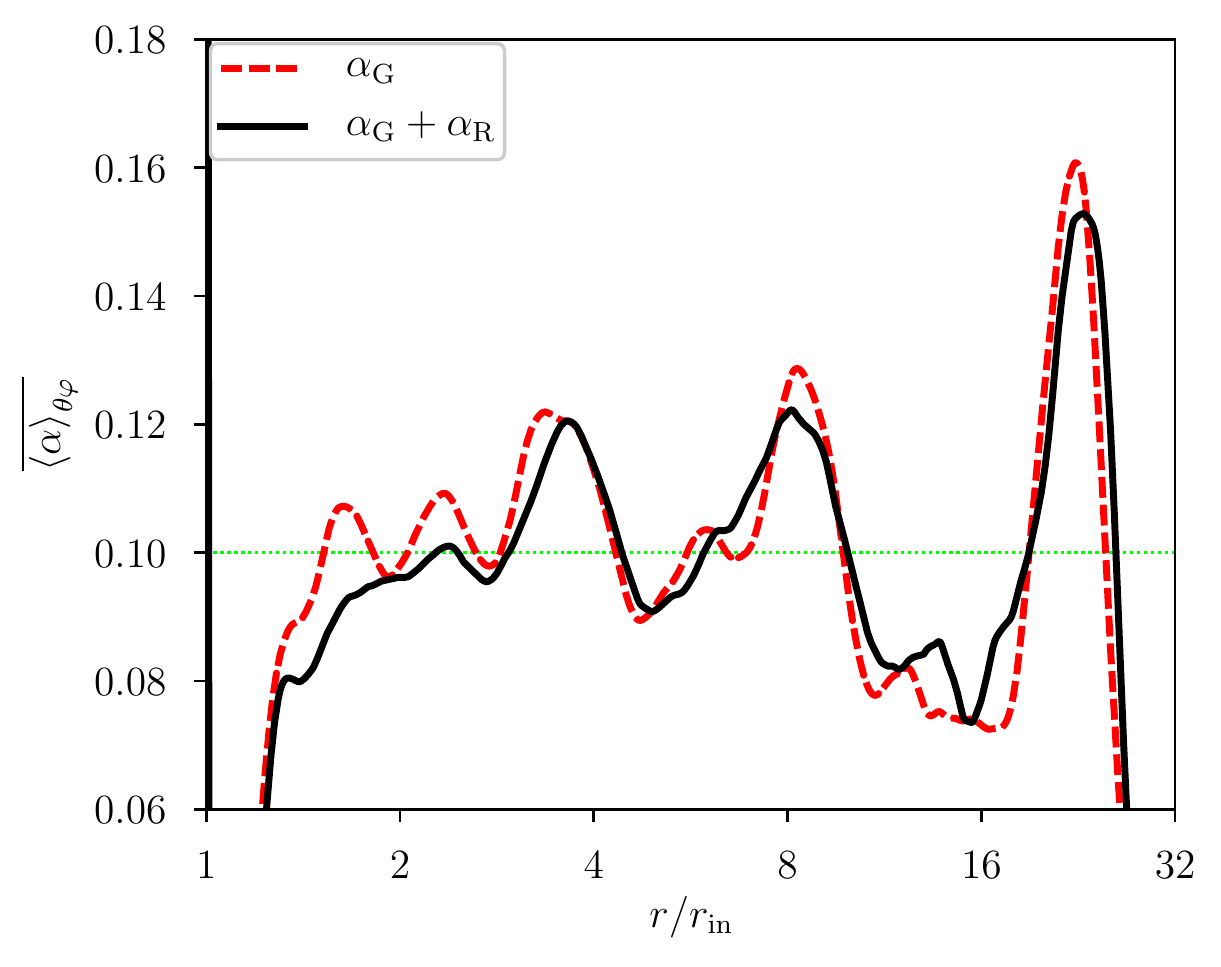}
  \caption{Radial profiles of the normalized total stress (solid black) and gravitational stress alone (dashed red) after spatial and time averaging over $181 t\lin$ in run M3B10. The dotted horizontal line marks the expected value $\alpha_{\mathrm{LTE}}$ assuming a Keplerian shear rate in Eq. \eqref{eqn:alpha0}.}
  \label{fig:m3b10_avg_alpha}
\end{figure}

We show in Fig. \ref{fig:m3b10_avg_alpha} the time-averaged radial profiles of stress in run M3B10. The total stress is positive and oscillates by less than thirty per cent of its mean value. The mean value $\overline{\brac{\alpha}_{r,\theta,\varphi}} = 0.10\pm 0.01$ is compatible with the predicted $\alpha_{\mathrm{LTE}}=0.1$ assuming a Keplerian shear rate $\dd\!\log \Omega / \dd\!\log R = -3/2$. In comparison, the hydrodynamic stress oscillates around zero with small amplitudes: $\alpha_{\mathrm{R}} \lesssim 10^{-1} \alpha_{\mathrm{LTE}}$. It is therefore the stress induced by fluctuations of the gravitational potential that predominantly transports angular momentum. This is consistent with previous global simulations \citep{lodato04,boley06b,forgan11}, although the exact contribution of $\alpha_{\mathrm{R}}$ is delicate to measure because of the large velocity deviations relative to their azimuthal average \citep{michael12}. We believe that our grid resolution is insufficient to capture small-scale parametric instabilities, which could invigorate the Reynolds stress up to $\alpha_{\mathrm{R}} \approx \alpha_{\mathrm{G}}/2$ at most \citep{riols17,booth19}. 

In principle, angular momentum could also be transported vertically. The fluctuations of the gravitational potential have the same phase as the density fluctuations, so the vertical gradient $\partial_z \Phi_{\mathrm{disk}}$ has a turning point above density extrema whereas the azimuthal gradient $\partial_{\varphi}\Phi_{\mathrm{disk}}$ changes sign. The $z \varphi$ component of the gravitational stress tensor therefore vanishes upon azimuthal averaging on the $\theta$ boundaries. In other words, the upper/lower boundaries exert no torque on the disk, so mass accretion is determined by the radial flux of angular momentum ($\alpha$) alone.

Because $\alpha$ vanishes at the inner radius, the inner radial boundary effectively imposes a stress-free condition. The steep increase of $\alpha$ inside $r/r\lin \lesssim 1.4$ drives a net inward mass flux in this region and it is responsible for most of the mass losses over time. In the rest of the domain, a constant $\alpha$ means that the angular momentum flux decreases radially as fast as the initially prescribed pressure profile $P \propto R^{-4}$, driving a net outward mass flux in agreement with the disk spreading shown in Fig. \ref{fig:m3b10_map_dn} \citep[see equation 28 of][]{balbuspap99}. We stress that the redistribution of mass does not represent a steady decretion disk. Only local thermal balance is satisfied over the timescales of our simulation, while the global mass profile keeps evolving.

While the time and azimuthally averaged stress coefficient $\alpha$ matches the expected $\alpha_{\mathrm{LTE}}$ and can be related to the average mass accretion rate, the root-mean-square dispersion of the stress reaches several $10^2 \alpha_{\mathrm{LTE}}$ at any given radius and time. At any time in the equatorial plane, we find large areas spanning $\delta \varphi \gtrsim \uppi/4$ and $\delta r / h \gtrsim 6$ exhibiting a negative $\mathrm{G}_{R\varphi}<0$ stress. It is remarkable that the stress fluctuations reduce precisely to $\alpha_{\mathrm{LTE}}$ upon averaging, but it is also expected since a viscous $\alpha$ disk theory should apply (only) in an average sense to quasi-steady disks \citep{balbuspap99}. In regards to the dynamics of objects embedded in the disk, they may actually feel a nonvanishing net torque upon orbit-averaging and migrate in a way that viscous disks models cannot account for \citep{baruteau11,michael11,stamatellos15}.

\subsection{Velocity dispersion} \label{sec:veldisp}

\begin{figure}
  \centering
  \includegraphics[width=\columnwidth]{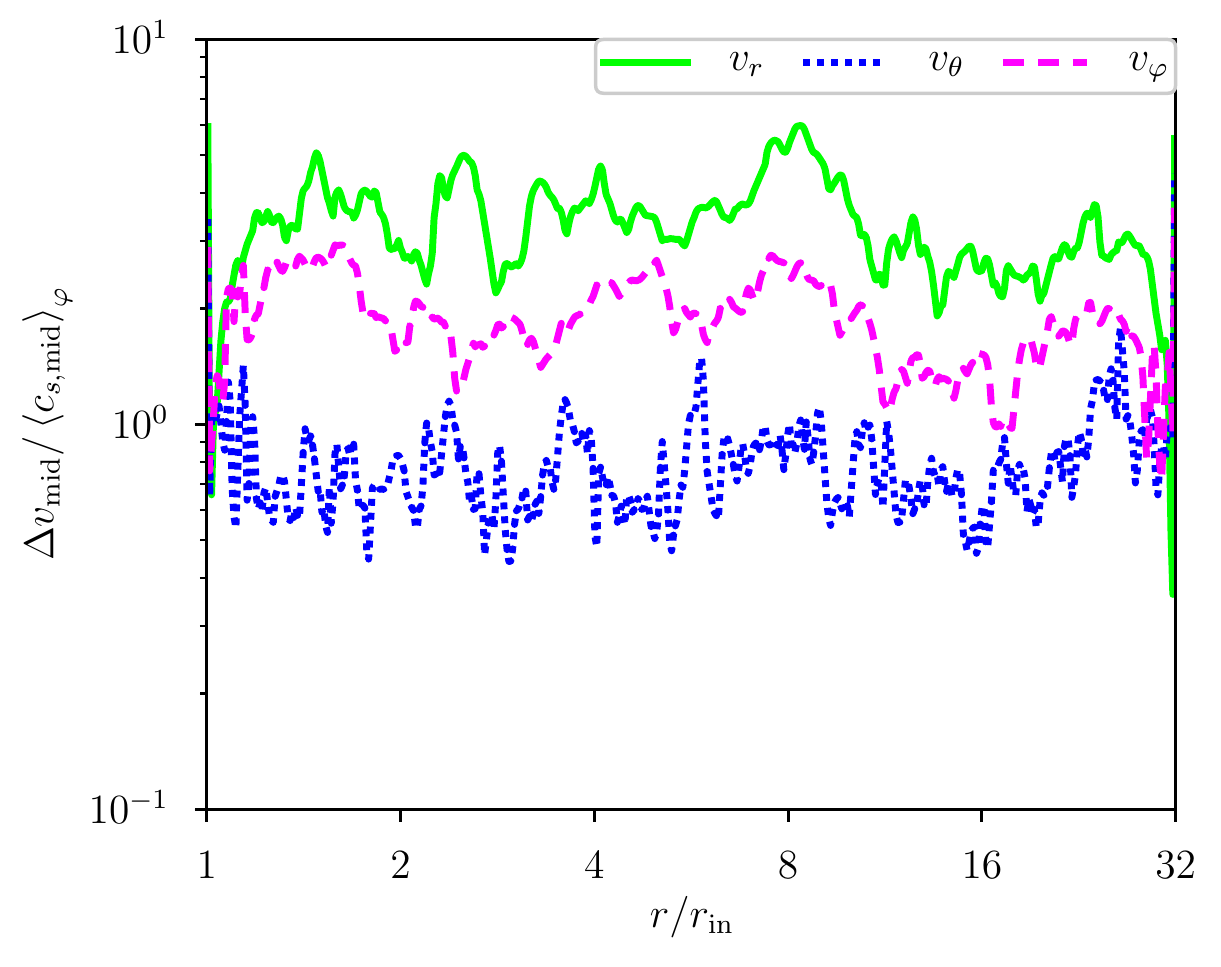}
  \caption{Radial profiles of the peak-to-peak range of the three velocity components in the midplane, relative to the azimuthally averaged midplane sound speed at time $500 t\lin$ in run M3B10.}
  \label{fig:m3b10_EQl_vcs_ptp}
\end{figure}

A common extrapolation of viscous disk theory is to connect the Mach number $\mathcal{M}$ of turbulent velocity fluctuations with the stress parameter $\alpha \sim \mathcal{M}^2$. This relation was used to extract a stress parameter $\alpha$ from turbulent line broadening \citep{flaherty15,flaherty18} or from the vertical settling of dust grains against turbulent diffusion \citep{dubrulle95,pinte16}. It was also used to constrain the largest grain sizes surviving collisional fragmentation for a prescribed value of $\alpha$ \citep{birnstiel09,birnstiel10}. However, the inequality $\alpha \leq \mathcal{M}^2$ holds for a purely hydrodynamic stress $\mathrm{R}_{R\varphi}$ and does not necessarily apply if the stress is dominated by the gravitational contribution $\mathrm{G}_{R\varphi}$. We verify the validity of this relation by examining the peak-to-peak range of all three velocity components in the disk midplane as a function of radius, as shown in Fig. \ref{fig:m3b10_EQl_vcs_ptp}.

The velocity dispersion clearly satisfies the above inequality: the squared turbulent Mach number $\mathcal{M}^2$ is two orders of magnitude larger than the average stress coefficient $\alpha$, so in fact $\alpha \ll \mathcal{M}^2$. This remains true by a factor $\sim 10$ when taking standard deviations over $\varphi$ of each velocity component instead of their peak-to-peak ranges. Considering that the hydrodynamic stress amounts to $\alpha_{\mathrm{R}} / \alpha \lesssim 10^{-1}$ of the total stress at most, we deduce that the $v_r$ and $v_{\varphi}$ velocity fluctuations are weakly correlated on average. The Reynolds stress tensor is essentially diagonal, so the transsonic velocity fluctuations mainly induce an effective pressure. The relation $\alpha_{\mathrm{R}} \approx \mathcal{M}^2$ would thus under-estimate the turbulent velocity dispersion by more than two orders of magnitude for a given $\alpha_{\mathrm{R}}$.

The extreme values reported in Fig. \ref{fig:m3b10_EQl_vcs_ptp} most likely correspond to the largest structures developing in the disk, namely, the spiral wakes. From the equatorial velocity components alone we estimate the typical velocity jump $\sqrt{\Delta v_r^2 + \Delta v_{\varphi}^2} \approx 4 \brac{c_s}_{\varphi}$ across the wakes, as previously estimated from Fig. \ref{fig:m3b10_EQm_vrcs}. Meanwhile, the vertical velocity fluctuations also reach a significant fraction of the sound speed in the midplane. These vertical motions could promote vertical mixing \citep{riols18} and alter the morphology and propagation of the spiral wakes, which we examine in the following section. 

\subsection{Spiral wakes} \label{sec:spirals}

In this section, we attempt to average the action of the small-scale turbulent fluctuations and wakes so as to examine their cumulative effect, and thus to better isolate the generic properties of large-scale spiral structures in our reference run M3B10. 

\subsubsection{Spiral pitch angle} \label{sec:openangle}

\begin{figure}
  \centering
  \includegraphics[width=\columnwidth]{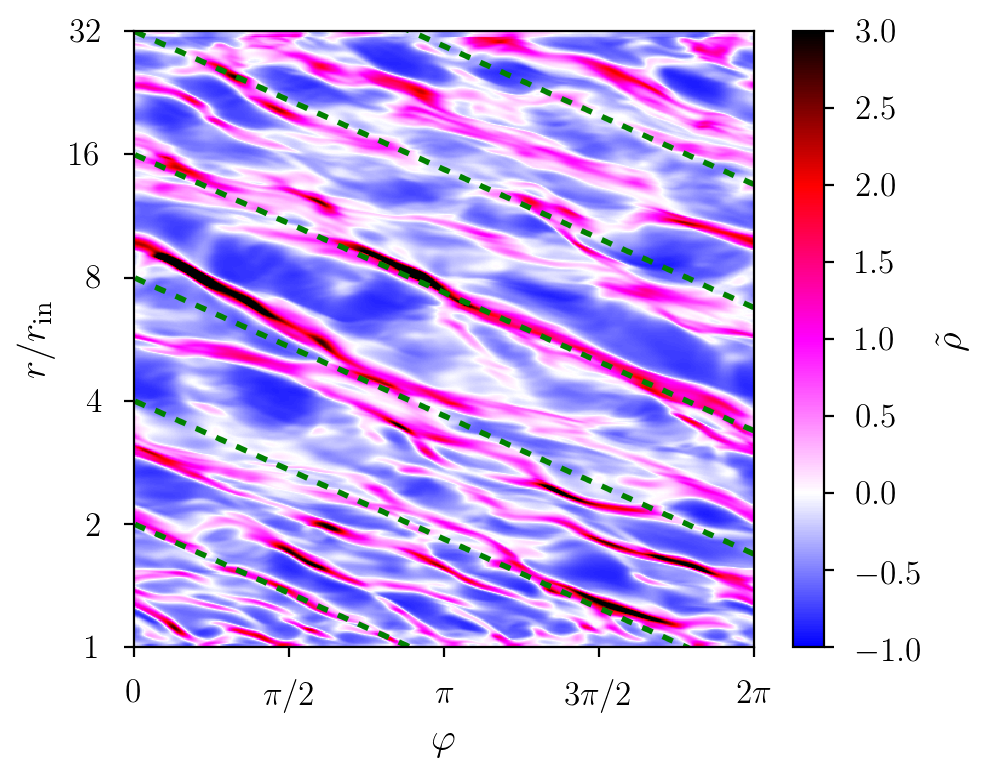}
  \caption{Vertically averaged density fluctuations $\tilde{\rho}$ at time $500 t\lin$ in run M3B10. The slope of the dashed lines refers to the estimated mean pitch angle $\tan\left(i\right) \approx 0.25$, or $i \approx 14^\circ$.}
  \label{fig:m3b10_EQt_ddn}
\end{figure}

We show in Fig. \ref{fig:m3b10_EQt_ddn} the map of the relative density fluctuations $\tilde{\rho}$ after vertical averaging at time $500 t\lin$ in run M3B10. The surface density fluctuates in the range $\tilde{\rho} \in \left[-0.9,4\right]$ relative to its azimuthal average at every radius. The density maxima in Fig. \ref{fig:m3b10_EQt_ddn} are located on the $v_r=0$ wake fronts visible in Fig. \ref{fig:m3b10_EQm_vrcs}. The morphology of these wakes can be parametrized locally by Eq. \eqref{eqn:spiralfront}, where the pitch angle $i$ is allowed to vary with radius and from one wake to another. For a pitch angle independent of $r$, this equation describes logarithmic spirals $\log\left(r/r_0\right) = -\tan \left(i\right) \left(\varphi-\varphi_0\right)$ and corresponds to straight oblique lines in Fig. \ref{fig:m3b10_EQt_ddn}. 

As is clear from Fig. \ref{fig:m3b10_EQt_ddn}, the wake fronts appear to share similar pitch angles over the entire disk. As all the dimensionless numbers ($Q$, $\beta$ and $h/R$) are roughly constant with radius, perhaps this was to be expected. In order to obtain a quantitative handle on these features we designed the following measure of the characteristic pitch angle. Instead of trying to fit lines over every visually identified wakes, as in the figure, we used the information contained in the entire plane by computing the 2D autocorrelation of the relative density fluctuations:
\begin{align}
    C\left[\tilde{\rho}\right]\left(\delta \log r, \delta \varphi\right) = \int\!\!&\!\int \tilde{\rho} \left(\log r, \varphi \right) \\
    &\times \tilde{\rho}\left(\log r + \delta \log r, \varphi + \delta \varphi\right) \dd\!\log r \dd \varphi, \nonumber
\end{align}
which represents how values of $\tilde{\rho}$ between two points separated by $\left(\delta \log r, \delta \varphi \right)$ are correlated on average. It is maximal when $\left(\delta \log r, \delta \varphi \right)$ separates two points on density maxima (minima), either along one wake front or separating different fronts. It is minimal when $\left(\delta \log r, \delta \varphi \right)$ separates a density maximum from a density minimum on average, for which the values of $\tilde{\rho}$ have opposite signs. The extrema of the autocorrelation are thus distributed on a series of oblique bands in the $\left(\delta \log r,\delta \varphi\right)$ plane, whose slope is the desired pitch angle.

\begin{figure*}[t!]
  \centering
  \includegraphics[width=\textwidth]{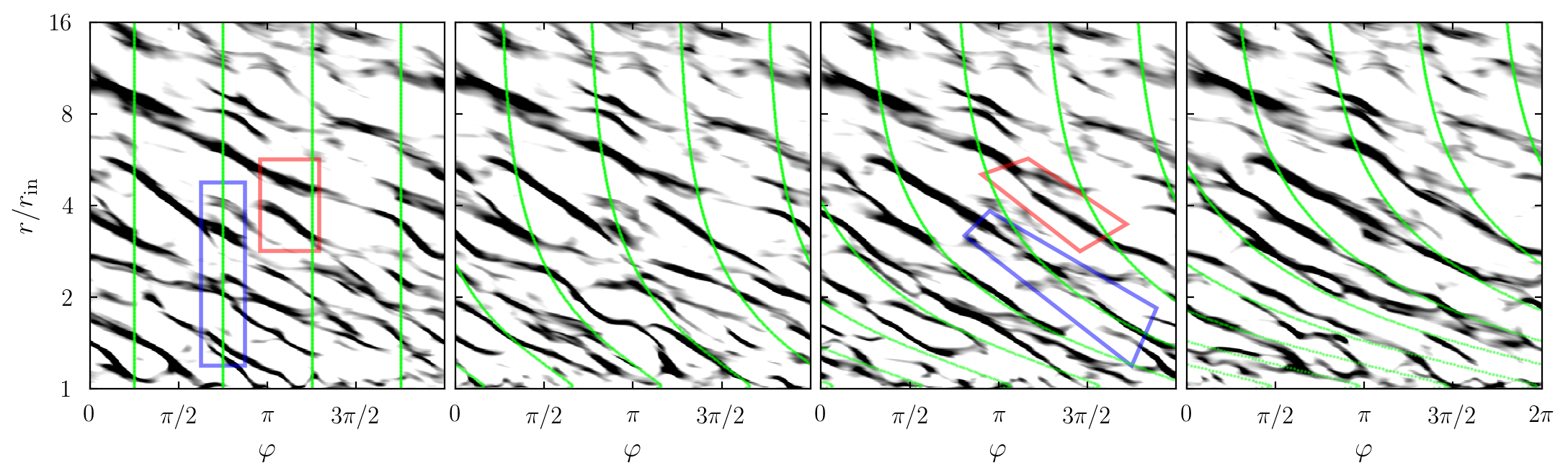}
  \caption{Relative density fluctuations at four consecutive times separated by $0.5 t\lin$ in the equatorial plane of run M3B10. The gray scale covers $\tilde{\rho} \in \left[0,1\right]$. The green curves are transported at the mean orbital velocity $\brac{v_{\varphi}}_{\rho}$ of the gas and serve as corotation references. The blue and red contours delimit initially disjoint wakes in the process of merging by the third snapshot.}
  \label{fig:m3b10_timestack}
\end{figure*}

We fitted a line passing through the origin $\left(\delta \log r, \delta \varphi\right) = \left(0,0\right)$ and maximized the integral of the autocorrelation:
\begin{equation}
  S\left(\theta\right) \equiv \sum_{\delta \varphi_k=-\uppi}^{\uppi} C\left[\tilde{\rho}\right]\left(\delta \log r = -\tan\left(\theta\right) \delta \varphi_k, \delta \varphi_k\right).
\end{equation}
The curvature of this integral near the optimal pitch angle was used to estimate an uncertainty $\delta i = \sqrt{S / \left(2 \dd^2 S / \dd \theta^2\right)}$. We repeated this procedure over fifteen simulation snapshots spanning $7.5 t\lin$ and assimilated the best fit values into a single estimate of the pitch angle and of its uncertainty. The dashed lines drawn in Fig. \ref{fig:m3b10_EQt_ddn} illustrate the estimated mean pitch angle $\tan\left(i\right) \approx 0.25$. It fits the slope of the wake crests and troughs on average over the whole plane, supporting our assumption of self-similarity. It is over five times larger than the expected $\tan\left(i\right) = \brac{h/R}_{\rho} \approx 0.05$ for tidally induced spiral density waves in non-self-gravitating razor-thin disks \citep{ogilvie02}.

The pitch angle relates the wavenumbers $\left(k,m\right)$ via Eq. \eqref{eqn:spiralfront}. In run M3B10 we evaluate $m\in\left\lbrace 2,4\right\rbrace$ for the dominant spirals, with no clear radial trend over time. We use this information in Sect. \ref{sec:spectrum} to connect back with the linear dispersion relation \eqref{eqn:disp2d}.

As all spirals share a similar pitch angle regardless of radius and time, they either propagate without being sheared out, or they emerge intermittently as locally unstable perturbations with the same narrow spectrum of wavenumbers. Should these be understood as standing waves or tightly wound traveling wave trains, their angular pattern speed would be independent of radius. To better understand the nature of these spiral structures we move on to examine their propagation in the disk plane.

\subsubsection{Equatorial motions} \label{sec:eqmotion}

As an overview, we show in Fig. \ref{fig:m3b10_timestack} the relative density fluctuations in four consecutive snapshots spanning two inner Keplerian periods in run M3B10. At all radii the spiral wakes follow the green curves, which are corotating with the gas. Examining smaller radii $r/r\lin\lesssim 6$, density peaks that are initially disjoint and sitting on the same vertical line eventually meet to form one larger canted wake, as highlighted by specimens in the colored boxes. Meanwhile, other wakes that are extended radially in the first snapshot are sheared apart into smaller disjoint fragments corotating with the flow. In other words, large-scale spiral arms only appear momentarily, when wakes at different radii and drifting at different velocities meet. It then follows that the logarithmic spiral fit illustrated in Fig. \ref{fig:m3b10_EQt_ddn} describes these individual wake fragments, since large-scale spiral arms only exist in a statistical sense, as a product of many locally corotating wakes sharing the same wavenumbers. 

\begin{figure}
  \centering
  \includegraphics[width=\columnwidth]{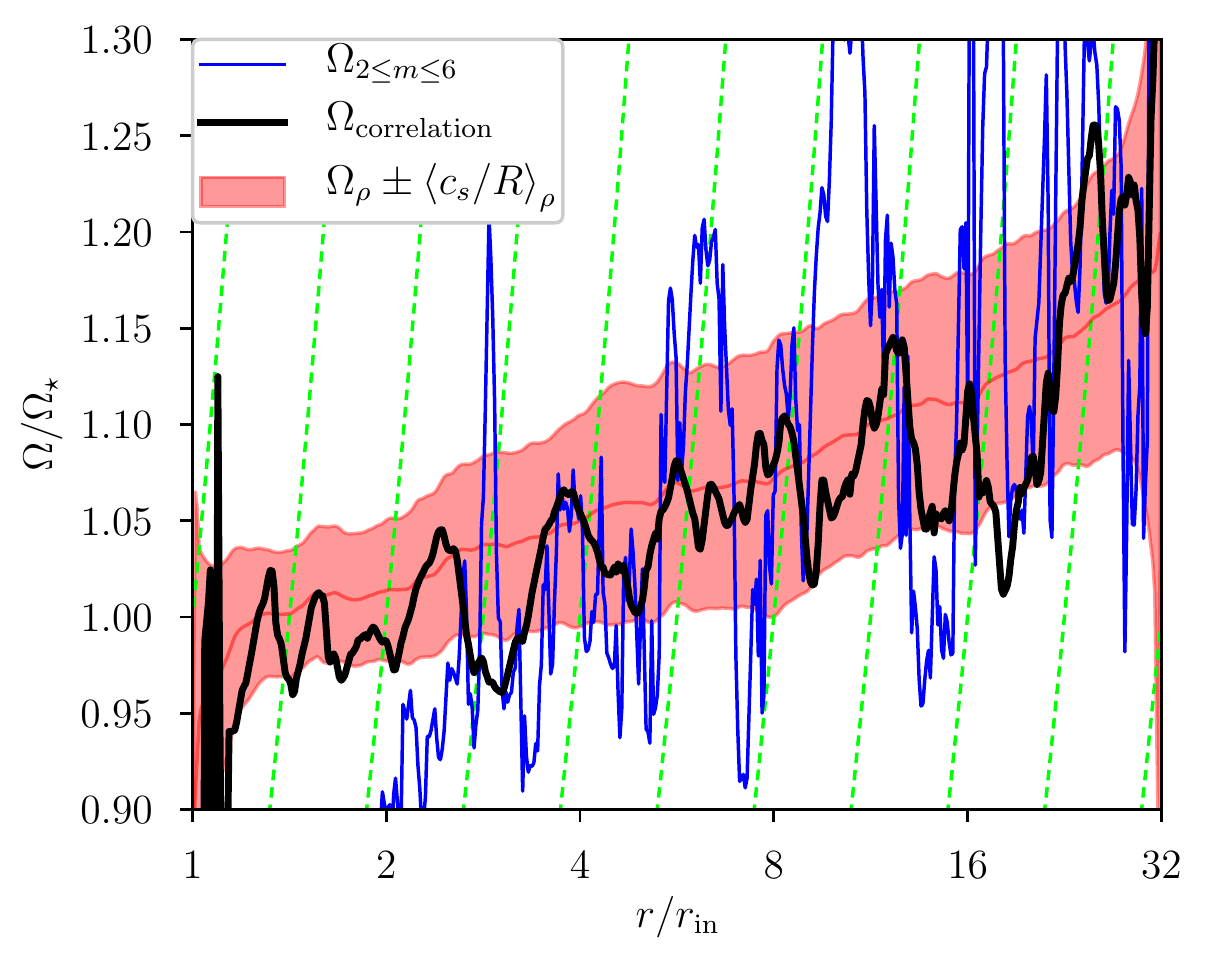}
  \caption{Radial profiles of angular velocities relative the Keplerian frequency $\Omega_{\star} \propto r^{-3/2}$ after averaging over $7.5 t\lin$ in run M3B10. The red area encompasses the bulk orbital velocity plus or minus the bulk sound speed. The thin blue curve is the spiral pattern speed estimated from the $2 \leq m \leq 6$ Fourier modes. The thick black curve is the pattern speed estimated by maximizing the time correlation \eqref{eqn:zomega}. The dashed green lines are contours of constant angular velocity.}
  \label{fig:m3b10_EQl_omg}
\end{figure}

To quantify how far the wakes propagate through the gas, we measure their angular pattern speed, assuming that it admits a representative value $\Omega_s$ at every radius. Our first measurement method uses azimuthal Fourier transforms of the gas surface density. At a given radius we extract the azimuthal phase of a Fourier component, compute the phase speed from consecutive times, and divide it by the azimuthal order $m$ to obtain a pattern speed. We computed the median pattern speed of the $2 \leq m \leq 6$ Fourier components in $15$ snapshots spanning $7.5 t\lin$. We denote the resulting pattern speed $\Omega_{2 \leq m \leq 6}$ and plot its radial profile in blue in Fig. \ref{fig:m3b10_EQl_omg}. This curve fluctuates around the gas orbital frequency $\Omega_{\rho} \equiv \brac{v_{\varphi}/R}_{\rho}$ (in red) with supersonic amplitudes on short length scales. It shows signs of coherent radial propagation only over narrow intervals where $\Omega_{2 \leq m \leq 6}$ roughly follows the dashed green lines of constant angular velocity (e.g., at $r/r\lin \in \left[7,10\right]$). Though somewhat noisy, this curve indicates that the disk splits into narrow strips of coherent wave activity, but that overall the pattern speeds track the orbital frequency.

Concerned with the robustness of this estimator against spatial discontinuities and fluctuations in time, we designed an alternative method to measure the spiral pattern speed. 
At a given radius, we assume that the wake pattern is transported in the $\varphi$ direction at a constant angular velocity independent of $\varphi$ and time. The angular velocity $\Omega_{\mathrm{correlation}}$ is estimated by maximizing the correlation $Z\left(\omega\right)$ between $N-1$ pairs of consecutive snapshots separated in time by $\delta t_n$: 
\begin{equation} \label{eqn:zomega}
Z \left( \omega \right) \equiv \sum_{n=1}^{N-1} \int_{0}^{2\uppi} f\left(\varphi, t\right) f\left(\varphi+\omega \,\delta t_n, t+\delta t_n\right)\dd \varphi,
\end{equation}
where we compressed the density fluctuations using 
\begin{equation}
f \equiv \tanh \left(\frac{\tilde{\rho}}{\std{\tilde{\rho}}}\right)
\end{equation}
to smooth the extrema and focus the optimization problem on well defined wake patterns. While the Fourier method projects the superposition of nonlinear spirals onto low-order smooth functions of $\varphi$, the correlation method is sensitive to the wake pattern as a whole and yields its mean angular velocity. The resulting profile of $\Omega_{\mathrm{correlation}}$ is drawn in black in Fig. \ref{fig:m3b10_EQl_omg}. The deviations from corotation are on the order of the bulk sound speed and biased toward negative values on average:
\begin{equation} \label{eqn:corotation}
  \frac{\Omega_{\mathrm{correlation}} - \Omega_{\rho}}{\Omega_{\rho}} = \left(-0.5 \pm 0.7\right) \brac{h/R}_{\rho} \approx 2.25 \times 10^{-2}.
\end{equation}
We conclude that the spiral patterns move at the gas orbital velocity to better than $\brac{h/R}_{\rho} \approx 0.05$ accuracy on average, meaning that they typically propagate at subsonic velocities. Intervals of roughly constant $\Omega_{\mathrm{correlation}}$ are present but they are still narrow and do not necessarily coincide with those of $\Omega_{2 \leq m \leq 6}$.

In contrast, \citet{cossins09} reported significantly larger values of $\xi \equiv \vrac{\left(\Omega_s - \Omega\right)/\,\Omega} \gtrsim 0.08$ at lower disk masses $M\leq 0.125$ ($h/R \lesssim 0.03$), and found that $\xi$ increases with $M$. At larger masses $0.5 \leq M \leq 1.5$, \citet{forgan11} reported values as high as $0.5 \leq \xi \leq 1.5$ throughout the disk. This is problematic because WKBJ waves far from corotation should be stable when $Q\approx 1$, so the disk would nowhere be unstable. The tension presumably originates in their use of the dispersion relation \eqref{eqn:disp3d} to deduce $\vrac{\Omega_s-\Omega}$ from measured wavenumbers $\left(k,m\right)$, whereas we directly track the gas kinematics. 

Following the arguments of \citet{cossins09}, large values of $\xi$ would imply that a substantial fraction of the energy extracted by the instability is carried away by waves before thermalization. This picture assumes a genuine propagation of the spirals through the gas, which would appear as oblique lines of constant $\Omega_s$ in Fig. \ref{fig:m3b10_EQl_omg}. Our profiles of $\Omega_s$ are never constant over significant distances, so we are lead to believe that the spirals are neither global wave modes nor traveling WKBJ wave trains. The excitation of coherent waves might be facilitated if, unlike in our case, there existed a preferential forcing timescale \citep[e.g., in a circumbinary disk,][]{desai19} or corotation radius \citep{mejia05,boley06b}, which gravito-turbulence does not seem to produce by itself. To complete our measurements of local corotation, we move on to examine the wakes in their proper corotating frame.

\subsubsection{Wake proper motions} \label{sec:properwake}

\begin{figure}[t]
  \centering
  \includegraphics[width=\columnwidth]{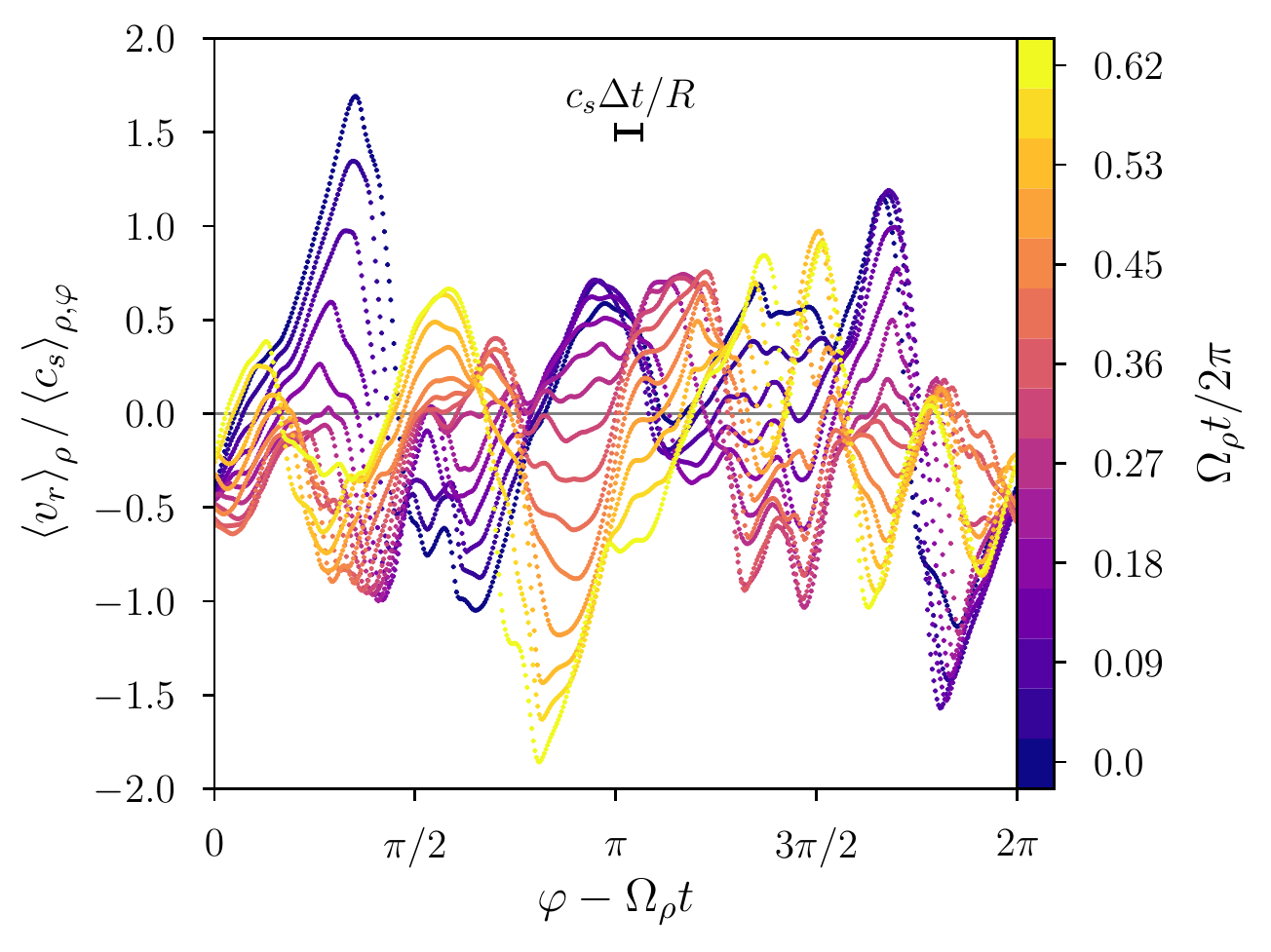}
  \caption{Radial velocity fluctuations in a frame corotating with the gas in run M3B10 at $r=5.2 r\lin$. The horizontal axis $\varphi-\Omega_{\rho} t$ corresponds to an azimuthal coordinate in the comoving frame. The vertical axis is the density-weighted mean radial velocity $\brac{v_r}_{\rho}$ relative to the average sound speed $\brac{c_s}_{\rho,\varphi}$. Different curves correspond to different times as color-coded in units of local orbital periods. The horizontal segment illustrates how far a signal would propagate in the $\varphi$ direction at the sound speed $\brac{c_s}_{\rho,\varphi}$ over this time interval.}
  \label{fig:m3b10_pattern_v1_5}
\end{figure}

If the spiral wakes were strictly corotating with the gas at every radius then they would quickly be sheared out by the disk's differential rotation. The most probable reason for all wakes to share similar pitch angles is that they form and vanish intermittently, and that we observe them only at their peak amplitude during a cycle of growth, swing and decay \citep{mamachag07,heinemann09,heinemann12}. To test this hypothesis, we select a radius in the disk and follow the flow at its local orbital velocity. Doing so, we can disentangle the proper evolution of a wake from its orbital motion.

Figure \ref{fig:m3b10_pattern_v1_5} shows the evolution of the mean radial velocity $\brac{v_r}_{\rho}$ relative to the azimuthally averaged bulk sound speed $\brac{c_s}_{\rho,\varphi} \equiv \brac{\brac{c_s}_{\rho}}_{\varphi}$ in a frame corotating with the gas at a radius $r/r\lin=5.2$. By looking at a single colored curve, one can inspect the instantaneous azimuthal structure of the flow from left to right. Reciprocally, by picking a specific absissa, one can inspect the evolution of the flow in time (colors) as seen in a frame corotating with the gas. In this frame, the propagation of a traveling wave in the $\varphi$ direction would appear as a steady horizontal translation of the pattern over time. 

All times considered, rather than azimuthally translating, the azimuthal profile of $\brac{v_r}_{\rho}$ oscillates around zero up to sonic amplitudes. The first and final times (dark and light, respectively) appear to be roughly anticorrelated along the horizontal axis: a maximum becomes a minimum over this time interval, and vice versa. This sign reversal occurs with moderate and irregular displacements in the horizontal direction, and suggests the presence of oscillation nodes reminiscent of a standing wave. The initial time displayed in Fig. \ref{fig:m3b10_pattern_v1_5} was chosen arbitrarily and we see similar reversals at different radii and in all simulations, pointing to a generic oscillatory phenomenon. We conclude that the spiral patterns are indeed corotating with the gas and do not exhibit the features of traveling wave trains. Since the compressive $\brac{v_r}_{\rho}$ motions are responsible for concentrating the gas into density wakes (see Figs. \ref{fig:m3b10_EQm_vrcs} \& \ref{fig:m3b10_EQt_ddn}), the reversal of $\brac{v_r}_{\rho}$ implies their destruction over orbital timescales. The wakes are therefore short-lived intermittent structures, corroborating the statistical nature of the large-scale spiral arms in our simulations.

\subsubsection{Instantaneous meridional structure} \label{sec:merryglyph}

\begin{figure}[t!]
  \centering
  \includegraphics[width=\columnwidth]{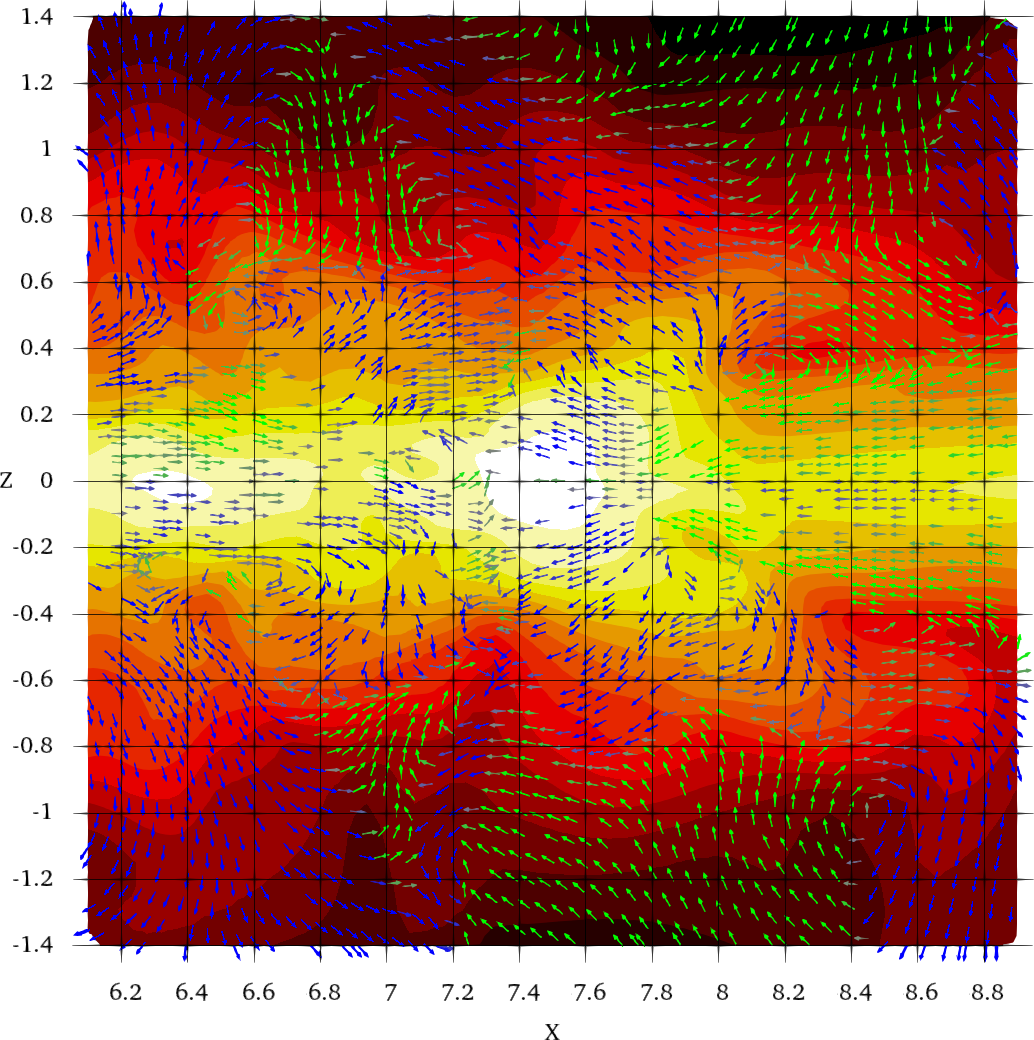}
  \caption{Instantaneous flow in a meridional slice at $\varphi = \uppi$, centered on a spiral wake at $R/r\lin \approx 7.5$ and time $500 t\lin$ in run M3B10. The gas density is represented by filled contours spanning four orders of magnitude. The orientation of the meridional gas velocity is represented with colored arrows, blue (resp. green) being oriented away (resp. toward) the disk midplane. At the wake location, the midplane pressure scale height $h_{\mathrm{mid}}/r\lin \approx 0.3$.}
  \label{fig:m3b10_glyphs}
\end{figure}

We have shown that the spiral wakes result from compressive motions in the disk midplane, we now examine the verticality of these motions. We cut a meridional slice through the disk at $\varphi=\uppi$, that is, roughly transversal to the wakes. We focus on the front visible at $R\approx 7.5 r\lin$ in Fig. \ref{fig:m3b10_EQt_ddn}, and represent the surrounding flow in Fig. \ref{fig:m3b10_glyphs}.

Away from the disk midplane ($\vrac{z} \gtrsim 0.6 r\lin \approx 2h$) the flow is organized in large scale rolls, as apparent from the horizontally alternating sign (blue and green colors) of the vertical velocity component. These rolls stretch up to the upper/lower $\theta$ boundaries and must therefore be controlled to some extent by our boundary conditions. However, their mirror symmetry with respect to $z=0$ points to a coherent behavior across the disk midplane, presumably correlated with the wake patterns.

Near the disk midplane ($\vrac{z} \lesssim 0.3 r\lin \approx h$) the flow is essentially horizontal and it converges toward the density maximum from both sides. The flow is smooth on scales $\lesssim h$, with no obvious discontinuities in the velocity and density fields, suggesting that shock heating is restricted to a fraction of the disk height. The flow in the vicinity of the wake is mainly oriented away from the midplane (blue color). This reorientation of the horizontal flow into vertical motions is smooth, with no indications of violent splashing nor shock bores \citep{boley06a}. While several smaller rolls can be identified within $2h$ of the midplane, it is unclear whether the associated circulations are pinned to the density maximum, as the flow is rather disorganized. We stress that the disk is strongly subadiabatic and thus not susceptible to convection. Nonetheless, roll motions can be induced by the baroclinic production of vorticity \citep{riols18}, and this is possibly what we are observing here. The reason that roll motions do not extend as far vertically as shown in \citet{riols18} may be attributed to the sheer strength of the stable stratification for the numerical set-ups we consider.

\subsubsection{Mean meridional structure} \label{sec:merry}

\begin{figure}[t!]
  \centering
  \includegraphics[width=\columnwidth]{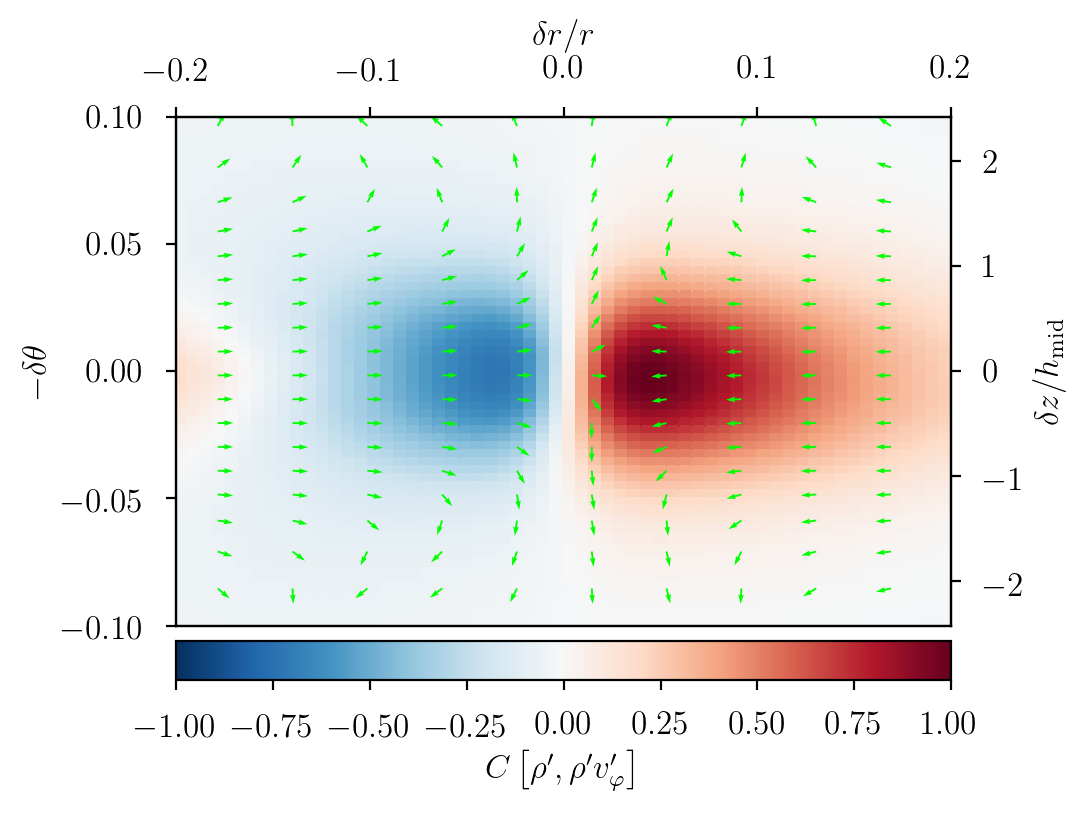}
  \caption{Mean meridional cross-correlation between the density fluctuations $\rho'$ and the momentum fluctuations $\rho' \bm{v}'$. The arrows give the orientation of the meridional components of the flow while the blue-red colors map its azimuthal component with an arbitrary scale. The right axis indicates the corresponding height measured from the midplane in units of mean midplane pressure scales.}
  \label{fig:m3b10_Abe_cc_mpfull}
\end{figure}

The superposition of turbulent motions makes it difficult to isolate the flow structure induced by the density wakes alone, so we designed the following filtering procedure. We normalized the density at every radius by its average value at that radius: $\rho' = \rho / \brac{\rho}_{\theta,\varphi}$. We also normalized the velocity fluctuations by the bulk orbital velocity: $\bm{v}' = \left(\bm{v} - \brac{\bm{v}}_{\rho} \right) / \brac{v_{\varphi}}_{\rho}$. Next, we defined the meridional cross-correlation of two quantities $X$ and $Y$ at a given azimuthal angle $\varphi$ as:
\begin{align} \label{eqn:cross}
  \begin{split}
  C\left[X,Y\right]\left(\delta \log r, \delta \theta\right) &= \int\!\!\!\int X\left(\log r, \theta\right) \\
  &\times Y\left(\log r + \delta \log r, \theta + \delta \theta\right) \dd\!\log r \dd \theta.
  \end{split}
\end{align}
By choosing $X=\rho'$, only the regions of high density fluctuations such as spiral wakes contribute to the integral. We thus obtain the mean meridional distribution of $Y$ as seen from the location a typical density maximum. We took $Y = \rho' \bm{v}'$ as a measure of the momentum fluctuations, including a second $\rho'$ weight to filter out the large velocity fluctuations away from the disk midplane. Finally, we averaged this meridional cross-correlation over $\varphi$ and over fifteen snapshots spanning $7.5t\lin$.

The resulting cross-correlation map is shown in Fig. \ref{fig:m3b10_Abe_cc_mpfull}. The center $\left(\delta \log r, \delta \theta\right) = \left(0,0\right)$ represents the location of a typical wake at an arbitrary radius in the disk. Positive (resp. negative) horizontal coordinates $\delta r / r$ are located outside (resp. inside) of the wake. The vertical axis $-\delta \theta$ measures an angle from the midplane and spans roughly $\pm 2h_{\mathrm{mid}}$, so it does not capture the largest rolls visible in Fig. \ref{fig:m3b10_glyphs} at high altitudes. 

Near the equatorial plane ($\delta \theta = 0$), the flow converges horizontally toward the wake from both sides before being reoriented vertically away from the midplane. The vertical component $C\left[\rho',\rho' v_z'\right]$ of the cross-correlation is in fact significant inside $\vrac{\delta r / r} \lesssim 0.05$, so the orientation of the flow changes mainly within one pressure scale of the wake. Looking at the azimuthal component $C\left[\rho',\rho' v_{\varphi}'\right]$, it is negative inside ($\delta r < 0$) of the wake and positive outside. This can be interpreted as a consequence of angular momentum conservation during the radial motions of the gas. The absence of apparent backflow returning to the midplane from high altitudes is caused by the density weights incorporated in Eq. \eqref{eqn:cross}, biasing the integral to keep the structures closest to the wake. In summary, Figs. \ref{fig:m3b10_glyphs} \& \ref{fig:m3b10_Abe_cc_mpfull} illustrate the genuinely 3D nature of the flow around spiral wakes and their ability to promote vertical mixing even in global simulations with a steep entropy stratification.

\section{Varying the disk mass and cooling time} \label{sec:sample}

\begin{table*}
\caption{Simulation label, initial disk mass, cooling timescale, integration time, mean aspect ratio, mean dimensionless stress, mean spiral pitch angle and mean standard deviation of the gas surface density. The label of the reference simulation M3B10 is marked with a dagger.}
\label{tab:recap}
\centering          
\begin{tabular}{l l c c c c c c c}
\hline\hline
Label & $M$ & $\beta$ & $T / t\lin$ & $\brac{h/R}_{\rho}$ & $\alpha$ & $\tan \left( i \right)$ & $\std{\Sigma}/\Sigma$ \\
\hline
M1B10 & $1$ & $10$ & $300$ & $\left(1.0 \pm 0.1 \right)\times 10^{-1}$ & $\left(1.0\pm 0.2\right)\times 10^{-1}$ & $0.28 \pm 0.04$ & $0.63 \pm 0.04$ \\
M2B10 & $1/2$ & $10$ & $300$ & $\left(6.4 \pm 0.4 \right)\times 10^{-2}$ & $\left(1.0\pm 0.1\right)\times 10^{-1}$ & $0.26 \pm 0.02$ & $0.63 \pm 0.03$ \\
M3B10$^{\dagger}$ & $1/3$ & $10$ & $500$ & $\left(4.5 \pm 0.2\right)\times 10^{-2}$ & $\left(1.0\pm 0.1\right) \times 10^{-1}$ & $0.25 \pm 0.01$ & $0.65 \pm 0.03$ \\
M3B32 & $1/3$ & $10^{3/2}$ & $500$ & $\left(4.1 \pm 0.3 \right)\times 10^{-2}$ & $\left(3.2 \pm 0.2\right)\times 10^{-2}$ & $0.24 \pm 0.03$ & $0.34 \pm 0.02$ \\
M3B100 & $1/3$ & $10^{2}$ & $500$ & $\left(4.4 \pm 0.5 \right)\times 10^{-2}$ & $\left( 1.0 \pm 0.3 \right) \times 10^{-2}$ & $0.22 \pm 0.02$ & $0.21 \pm 0.01$ \\
M5B10 & $1/5$ & $10$ & $500$ & $\left(2.8 \pm 0.1 \right)\times 10^{-2}$ & $\left( 9.7 \pm 0.4 \right) \times 10^{-2}$ & $0.24 \pm 0.01$ & $0.64 \pm 0.02$ \\
M10B10 & $1/10$ & $10$ & $500$ & $\left(1.4 \pm 0.1 \right)\times 10^{-2}$ & $\left( 8.5 \pm 1.0 \right) \times 10^{-2}$ & $0.226 \pm 0.005$ & $0.63 \pm 0.02$ \\
\hline                  
\end{tabular}
\end{table*}

In this section we sample different dimensionless cooling times $\beta$ and disk masses $M$ relative to the central object. To control the disk mass, we rescale the gas surface density $\Sigma$ while keeping the radial dependence $\Sigma\left(R\right) \propto R^{-2}$. We also keep the initial Toomre number $Q \approx 1$ by varying the initial sound speed in the same proportion as $\Sigma$. As cooling brings the disk toward a gravitationally unstable state, we always find that the density-weighted mean $\brac{Q}_{\rho} \approx 1$ in the turbulent phase. All simulations are integrated over more than $300 t\lin$ to characterize quasi-steady states. Since $\Sigma$ evolves slowly over the duration of a simulation, the density-weighted mean sound speed remains close to its initial profile and different disk masses correspond to different aspect ratios $\brac{h/R}_{\rho} \simeq \uppi \Sigma R^2 / m_{\star}$. For example, we verified that $\uppi \Sigma R^2 / m_{\star} = (1.09 \pm 0.05)\brac{h/r}_{\rho}$ in run M3B10. The simulations are listed in Table \ref{tab:recap} with their main parameters and diagnostics. 

\subsection{Low-mass disk} \label{sec:m1/10}

We start with run M10B10, which has the lowest mass considered, $M=1/10$ and the same cooling timescale $\beta=10$ as previously. For comparison, a global SPH simulation with the same parameters was presented by \citet{cossins09}, albeit with a different slope for the surface density profile $\Sigma \propto R^{-3/2}$. \citet{lodato04} also presented a simulation with $M=1/10$ but with a surface density $\Sigma \propto R^{-1}$ and a shorter cooling timescale $\beta=7.5$.

\subsubsection{Overview} \label{sec:lowmassz}

\begin{figure}
  \centering
  \includegraphics[width=\columnwidth]{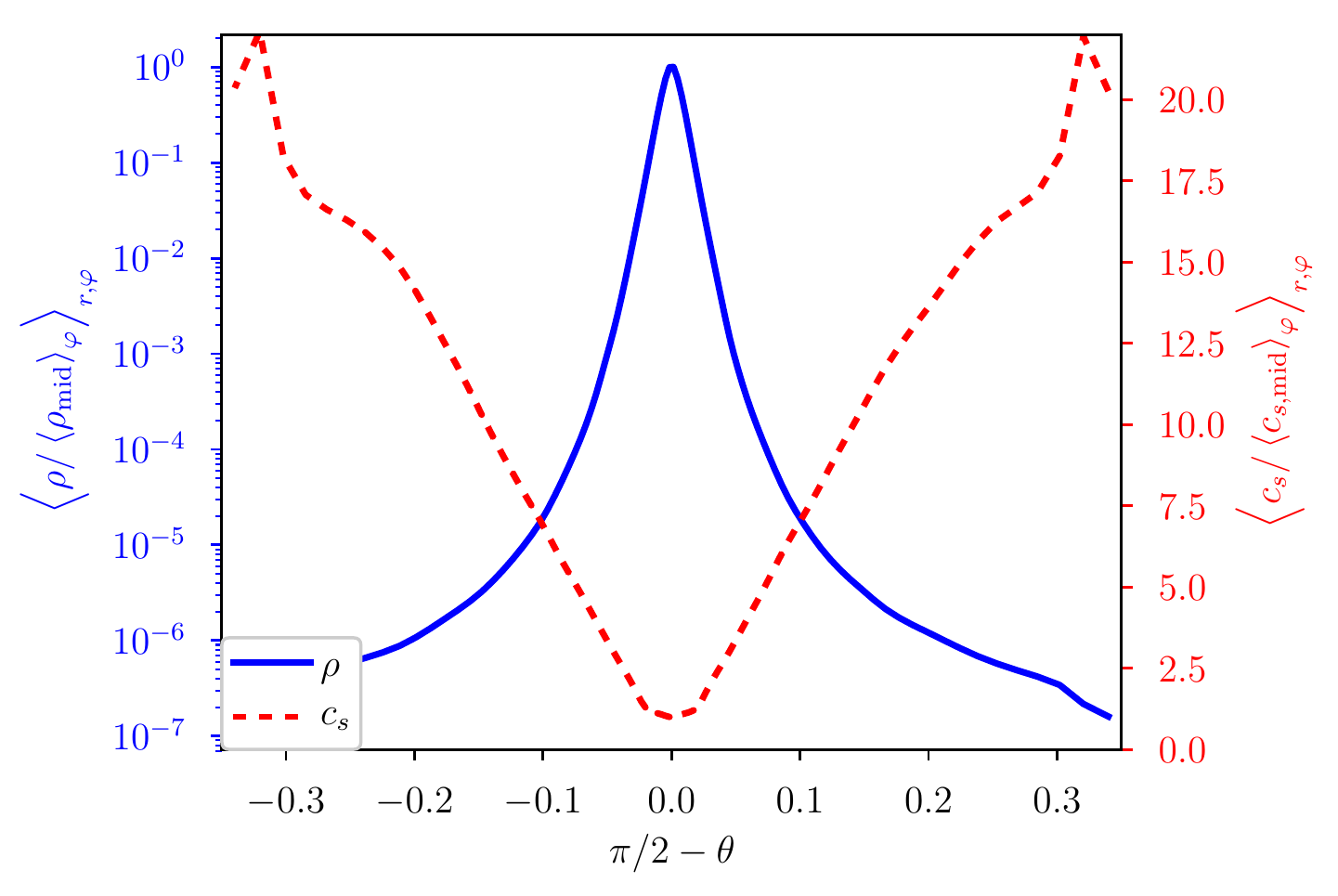}
  \caption{Same as Fig. \ref{fig:m3b10_COl_rho} at time $500 t\lin$ in the least massive case M10B10.}
  \label{fig:m10b10_COl_rho}
\end{figure}

\begin{figure}
  \centering
  \includegraphics[width=\columnwidth]{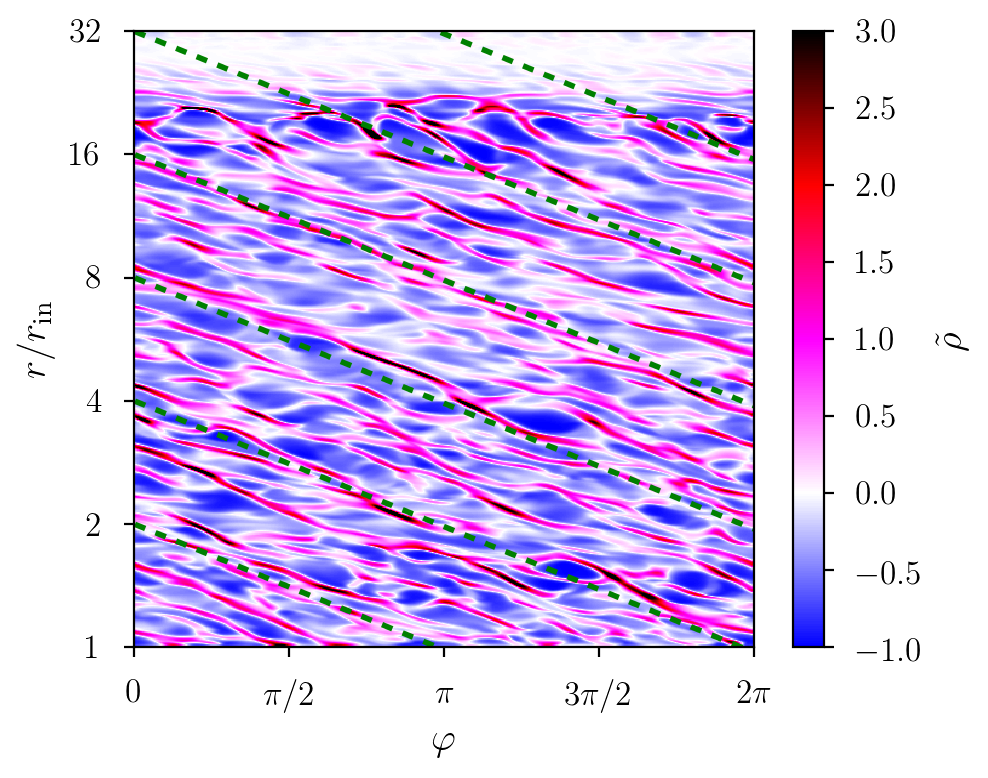}
  \caption{Same as Fig. \ref{fig:m3b10_EQt_ddn} in the least massive case M10B10. The dashed lines refer to a pitch angle of $\tan\left(i\right) \approx 0.23$, or $i \approx 13^\circ$.}
  \label{fig:m10b10_EQt_ddn}
\end{figure}

The disk of run M10B10 loses less than seven per cent of its mass over $500 t\lin$. After the initial transient, the Toomre number settles at $\brac{Q}_{\rho} \approx 1$ with fluctuations less than ten per cent across the disk. Given the smaller surface density $\Sigma$, marginal stability is attained for a smaller aspect ratio $\brac{h/R}_{\rho} \approx 1.4 \times 10^{-2}$. With less than five grid cells per midplane scale height, the vertical structure of the disk is only marginally resolved in this case.

We show the mean vertical structure of this disk at time $500 t\lin$ in Fig. \ref{fig:m10b10_COl_rho}. The gas density profile is strongly peaked in the midplane and nearly flat at high latitudes. The sound speed increases with height by up to a factor twenty relative to its midplane value. The temperature high above the disk is in fact comparable to that of run M3B10, confirming that this region is predominantly supported by pressure against the gravity of the central object both in the vertical and radial directions. 

\subsubsection{Spiral patterns} \label{sec:lowmasseq}

We show in Fig. \ref{fig:m10b10_EQt_ddn} the relative density fluctuations in the equatorial plane of run M10B10 at time $500 t\lin$. Spiral wakes fill the disk up to $r/r\lin \lesssim 22$ while turbulence is still developing in the outermost regions. The peak-to-peak equatorial velocity dispersion $\sqrt{\Delta v_r^2 + \Delta v_{\varphi}^2} \approx 5 \brac{c_s}_{\varphi}$ amounts to a similar turbulent Mach number as in run M3B10. Accordingly, the density fluctuations reach similar amplitudes as in run M3B10 with a standard deviation $\std{\Sigma}/\Sigma \approx 0.63$ (see Table \ref{tab:recap}).

Compared to run M3B10, the wake fronts are narrower and closer to each other, with a typical azimuthal orders $m \approx 8$. We employ the method described in Sect. \ref{sec:openangle} to measure the mean spiral pitch angle, using $21$ snapshots spanning $10 t\lin$. We find that $\tan\left(i\right) \approx 0.23$, slightly smaller than in run M3B10. This mean pitch angle is illustrated by the dashed oblique lines in Fig. \ref{fig:m10b10_EQt_ddn}, which fit the mean slope of the density wakes at all radii. 

Using the procedure described in Sect. \ref{sec:eqmotion} over the same $21$ snapshots, we measure the mean angular velocity of the spiral density patterns. In this case, the deviations from corotation amount to
\begin{equation} \label{eqn:m10corotation}
  \frac{\Omega_{\mathrm{correlation}} - \Omega_{\rho}}{\Omega_{\rho}} = \left(-0.4 \pm 1.0\right) \brac{h/R}_{\rho}  \approx 5.6\times 10^{-3}
\end{equation}
after radial averaging, smaller than in run M3B10.

Moving to a corotating frame as in Sect. \ref{sec:properwake}, we retrieve standing oscillations of the radial velocity over orbital timescales, such as appear in Fig. \ref{fig:m3b10_pattern_v1_5}, with little evidence for the propagation of the wake patterns. As in run M3B10, we conclude that the spiral patterns are nearly corotating with the gas and that they are destroyed and reborn before they can travel over significant radial distances.

\subsection{High-mass disk} \label{sec:m1}

\begin{figure}
  \centering
  \includegraphics[width=\columnwidth]{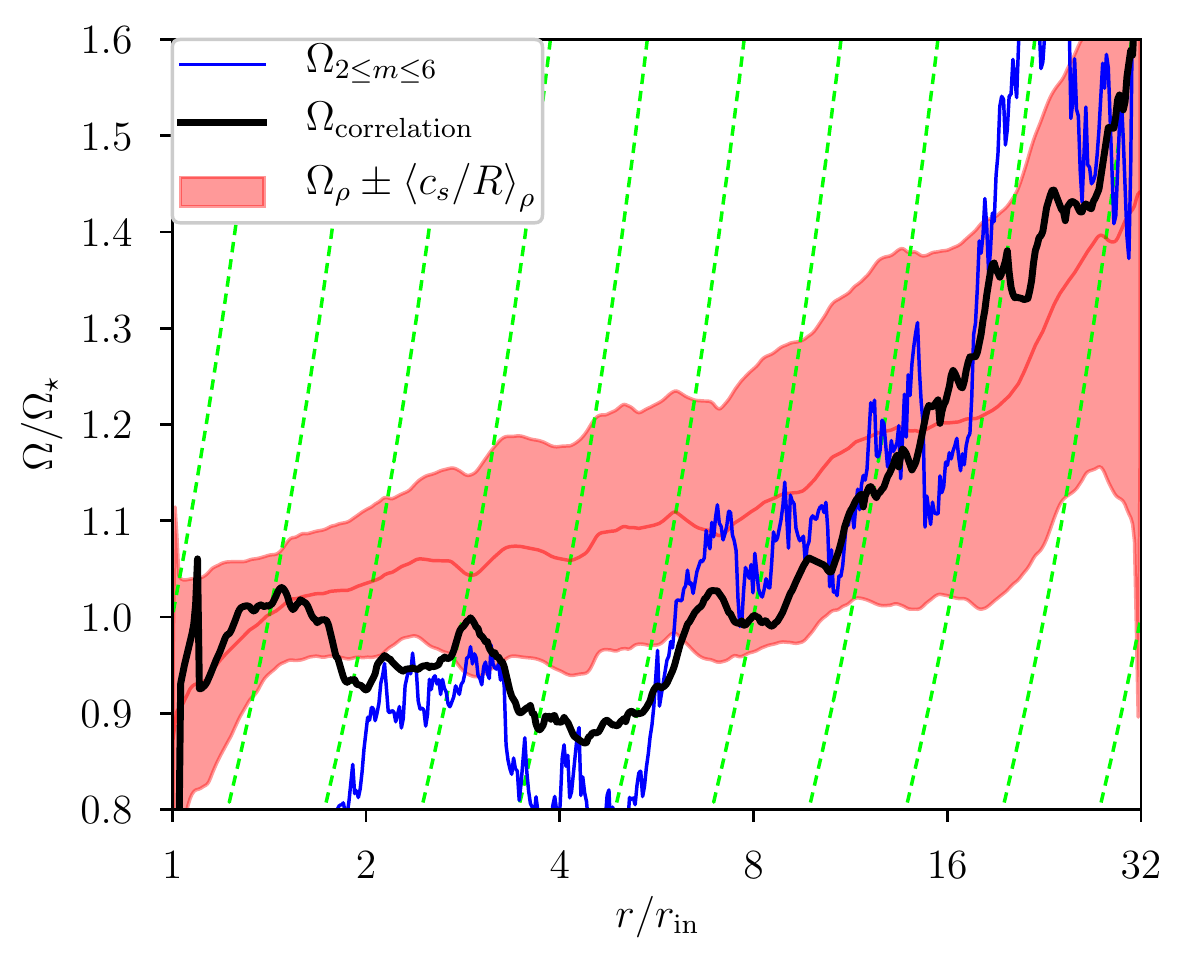}
  \caption{Same as Fig. \ref{fig:m3b10_EQl_omg} after averaging over $41$ snapshots spanning $20 t\lin$ in the most massive case M1B10.}
  \label{fig:m1b10_EQl_omg}
\end{figure}

\begin{figure}
  \centering
  \includegraphics[width=\columnwidth]{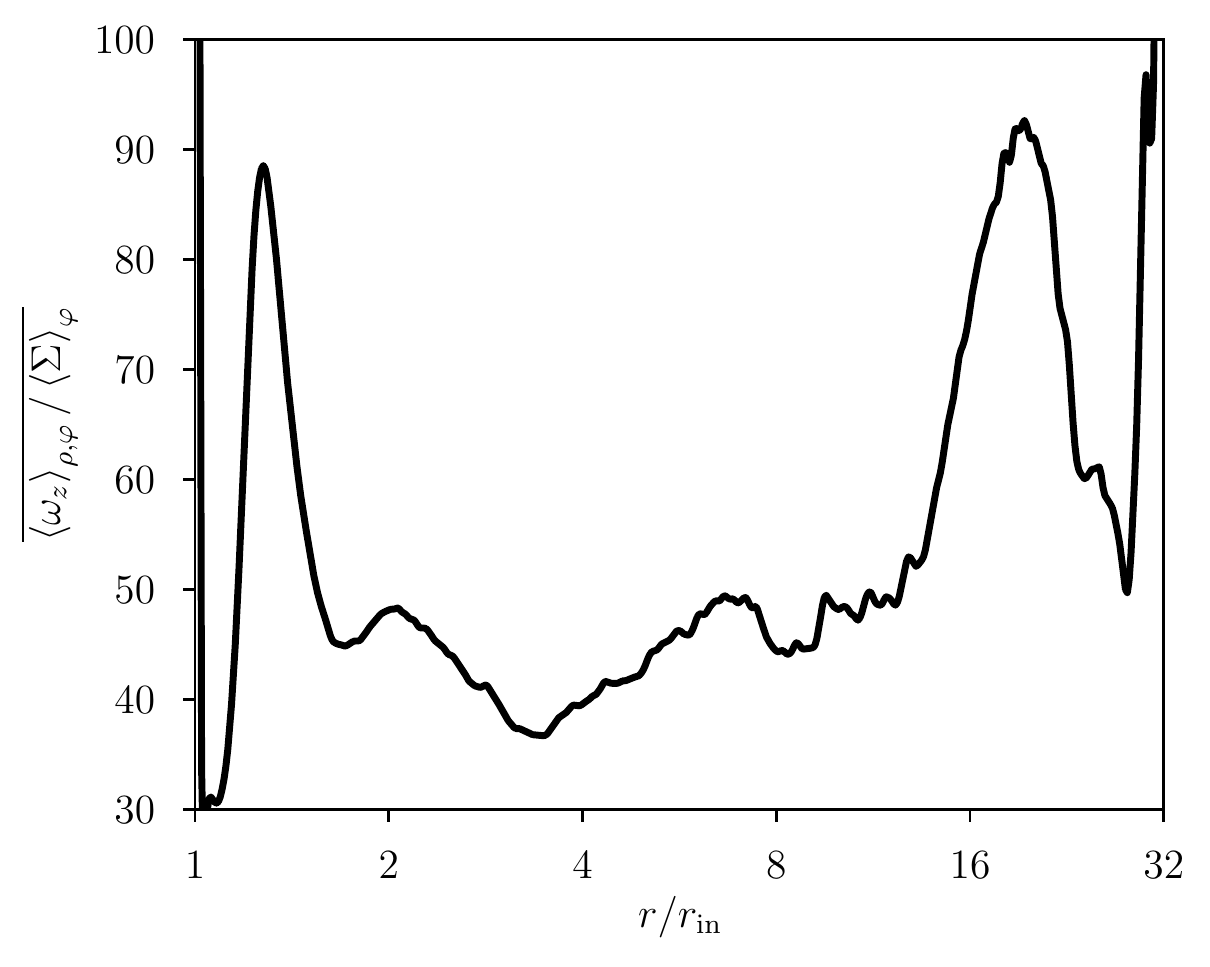}
  \caption{Radial profile of the vortensity in run M1B10. The maximum at $r/r\lin \approx 18$ coincides with the corotation radius in Fig. \ref{fig:m1b10_EQl_omg}.}
  \label{fig:m1b10_avg_pwz}
\end{figure}

Our most massive disk M1B10 has an initial mass equal to the central mass ($M=1$). Similarly massive disks were examined by \citet{lodato05} and \citet{forgan11}, again with different $\Sigma$ profiles. After the Toomre number has saturated to $\brac{Q}_{\rho} \approx 1$, the average thickness of the disk $\brac{h/R}_{\rho} \approx 0.1$. While the dominant spiral pattern has an azimuthal order $m=2$, one-armed spirals also occasionally appear. Such asymmetries displace the barycenter of the disk and should in principle cause the central object to react, an effect we neglect here but briefly consider in Appendix \ref{app:inertia}. Bearing this drawback in mind, we carry on to evaluate the pattern speed of the wakes in run M1B10.

We estimate the angular pattern speed $\Omega_s$ of the wakes via the two methods explained in Sect. \ref{sec:eqmotion}, using $41$ simulation snapshots over $20t\lin$. The corresponding pattern speed profiles are represented in Fig. \ref{fig:m1b10_EQl_omg} along with the bulk orbital velocity $\Omega_{\rho} \equiv \brac{v_{\varphi}/R}_{\rho}$ of the gas. The two methods are in relatively good agreement: the wake pattern speed is significantly slower than the gas in the inner $r/r\lin \in \left[2,8\right]$, corotating with the gas at $r/r\lin \approx 16$ and faster than the gas further out. Moving to a frame corotating with the gas at $r/r\lin \approx 5$ as in Sect. \ref{sec:properwake}, we do see a translation of the wake pattern as expected for traveling waves. These could be indications of spiral waves being excited near $r/r\lin \approx 16$ and propagating radially. 

After saturation of the GI, we observe a redistribution of the gas vorticity $\omega_z$ and surface density $\Sigma$ such that their ratio exhibit maxima at $r/r\lin \approx 1.4$ and $18$, see Fig. \ref{fig:m1b10_avg_pwz}. Such a vortensity profile only appears in run M1B10 and it can destabilize global waves even without invoking self-gravity \citep{papaloizou91}. Such instabilities should be attributed to the choice of initial and boundary conditions that lead to this vortensity profile, and could presumably be reproduced in lower-mass disks as well. In addition, we note that the $\Omega_{\mathrm{correlation}}$ and $\Omega_{2 \leq m \leq 6}$ curves only follow the lines of constant pattern speed near $r/r\lin \approx 16$ but clearly not over $r/r\lin \in \left[4,32\right]$. The deviations from corotation are saturated at sonic amplitudes, suggesting that waves cannot freely propagate due to shock dissipation. In summary, if global waves are excited in run M1B10, they may retain a local character by dissipating energy close to where it is extracted --- namely, near corotation.

\subsection{Spiral spectrum} \label{sec:spectrum}

From now on we consider our simulation set as a whole. In every simulation $m$ and $\tan\left(i\right)$ show no consistent variations with radius, so $k \propto R^{-1}$ on average and the spirals keep a logarithmic morphology. We list in Table \ref{tab:recap} the pitch angles measured via the method described in Sect. \ref{sec:openangle}. They differ from each other by twenty per cent overall, despite the cooling timescale $\beta$ and the disk mass $M$ varying by a factor ten. Considering only the simulations with a cooling time $\beta=10$, we can fit a linear dependence of the pitch angle on the aspect ratio: $\tan\left(i\right) \approx 0.222 + 0.602 \times \brac{h/R}_{\rho}$.

Using the measured $\tan\left(i\right)$ and the density-weighted average $k_0 \simeq \uppi G \Sigma / \brac{c_{s,\mathrm{iso}}}_{\rho}^2$, we estimate that 2D linear waves corotating with the gas at any location should possess $m_0=k_0 R \tan\left(i\right) \approx 6$ in run M3B10, $m_0\approx 9$ in run M5B10 and $m_0\approx 16$ in run M10B10. Instead, we identify spirals of order $m\in\left\lbrace 2,4 \right\rbrace$ in run M3B10, $m\in\left\lbrace 4,5 \right\rbrace$ in run M5B10 and $m\in\left\lbrace 7,9 \right\rbrace$ in run M10B10. Since the dominant azimuthal wavenumber is $m\approx m_0/2$, the radial wavenumber must be $k\approx k_0/2$ to account for the measured pitch angles. The discrepancy $k\neq k_0$ should come as no surprise given the several issues raised\footnote{From the vertical dilution of the potential alone, we would expect $k/k_0 \approx 0.466$ for the marginally stable modes of Eq. \eqref{eqn:disp3d} at $Q\approx 0.647$.} in Sect. \ref{sec:disprel}. On the other hand, the consistent scaling of $k$ with $k_0$ across the various simulations supports the basic thrust of the quasi-linear theory. 

The ratio of $k/k_0$ obtained by \citet{cossins09} can be deduced from their figure 13. They show that $m\left(\Omega_s-\Omega\right) / k c_s \approx 1$ regardless of disk mass and cooling time when injecting the measured mean wavenumbers $\left(k,m\right)$ at each radius in the WKBJ dispersion relation \eqref{eqn:disp3d}. This implies that the first two terms on the right hand side of Eq. \eqref{eqn:disp3d} cancel each other, or equivalently $k=k_0$ when $Q=1$. Their dominant radial wavenumber is thus twice larger than ours. It is possible that the discrepancy issues from the different vertical structures and discretizations of the disk. In particular, in their SPH simulations the disk scale height $h \lesssim 0.03 R$ was resolved by roughly two smoothing lengths (see their figure B5). If the spirals observed by \citet{cossins09} were closely corotating with the gas --- as we found in all cases --- then the equality $k=k_0$ could indicate that vertical structures and motions were not fully resolved, so that the flow was effectively 2D. It is also possible that their mean wavenumber $k$ was biased to larger values by the skewed power spectrum which they used as an averaging weight (see their figure 10).

We may extend our results to parametrize the predominant wavenumbers $\left(k,m\right)$ and describe the spiral structure analytically. For the azimuthal wavenumber:
\begin{equation} \label{eqn:empim}
    m = k R \tan\left(i\right) \approx 0.12 \left( h / R \right)^{-1},
\end{equation}
with the limit $m=1$ of one-armed spirals requiring a proper account of the reaction of the central object \citep{adams89}. The morphology of a spiral arm may then be described by
\begin{equation} \label{eqn:empik}
    \vrac{ \frac{\dd \varphi}{\dd R} } = \frac{k}{m} \approx \frac{1}{2} \frac{\kappa^2}{m\uppi G \Sigma},
\end{equation}
which is meaningful only on short radial intervals ($m$ must not change) and in a statistical sense given the local and transient nature of the spiral wakes. For our choice of $\Sigma$ profile, the right hand side of Eq. \eqref{eqn:empik} is $\propto R^{-1}$ and, on integrating, we obtain the logarithmic spiral structure seen throughout all our simulations. As mentioned, the crude relations \eqref{eqn:empim} \& \eqref{eqn:empik} may be sensitive to the detailed stratification of the disk and we caution against using them to quantitatively identify self-gravitating spirals in observations until this issue is settled.

\subsection{Density contrast} \label{sec:densdisp}

As is apparent in Figs. \ref{fig:m3b10_EQt_ddn} \& \ref{fig:m10b10_EQt_ddn} the spiral wakes are narrow and separated by large areas of low density. At any given radius, the maximum surface density fluctuation is approximately five times larger than the standard deviation $\std{\Sigma}$ at that radius in run M3B10, six times in run M3B32 and seven times in run M3B100. In comparison, this ratio would be $\sqrt{2} \approx 1.4$ for linear waves with a sinusoidal behavior. Despite nonlinearities, there exist a simple scaling between the density dispersion $\delta \Sigma$ and the cooling timescale: $\delta \Sigma / \Sigma \propto \beta^{-1/2}$ \citep{cossins09}. This relation could be used to constrain the observability of spirals in self-gravitating disks \citep{hall16,cadman20}.

We find that the formula $\std{\Sigma}/\Sigma = 2/\sqrt{\beta}$ is compatible with all the standard deviations gathered in Table \ref{tab:recap}. It is however twice larger than the density dispersion measured by \citet{cossins09} at similar aspect ratios $h/R \lesssim 0.03$. In Appendix \ref{app:inertia} we show that the density contrast does not change with the grid resolution, as reported by \citet[][see their figure A1]{cossins09}. If the discrepancy is not a direct consequence of our different spatial discretizations, then it may result from the different thermal stratifications and vertical motions in the two codes. 

\subsection{Angular momentum transport}

From the requirement of local thermal balance we expect the dimensionless stress $\alpha$ to take the constant value $\alpha_{\mathrm{LTE}}$ defined by Eq. \eqref{eqn:alpha0} at all radii. The time and radially averaged values of $\alpha$ gathered in Table \ref{tab:recap} are all compatible with $\alpha_{\mathrm{LTE}}$ to ten per cent accuracy. The largest deviation from $\alpha_{\mathrm{LTE}}$ is found in the least massive case M10B10. There, the gravitational stress $\alpha_{\mathrm{G}} \approx 1.2 \times 10^{-1}$ is larger than $\alpha_{\mathrm{LTE}} = 10^{-1}$, so the inequality $\alpha < \alpha_{\mathrm{LTE}}$ comes from a negative contribution of the average Reynolds stress. The results presented in Sect. \ref{sec:m1/10} support the idea of local thermal equilibrium in run M10B10, so this slight departure $\alpha \neq \alpha_{\mathrm{LTE}}$ likely comes from the marginal resolution of the disk in the vertical dimension \citep{michael12}. Overall, the agreement between the measured $\alpha$ and the theoretical $\alpha_{\mathrm{LTE}}$ supports the hypothesis of local energy conversion for all the disks considered.

\section{Summary} \label{sec:summary}

We simulated the global 3D hydrodynamics of gravitational instability with a shock-capturing grid-based code in self-similar models of self-gravitating disks. We considered disk masses in the range $M \in \left[1/10,1\right]$ relative to that of the central object, for which a Toomre number $Q=1$ corresponds to aspect ratios $h/R \in \left[0.01,0.1 \right]$. We prescribed a local cooling law with a characteristic timescale $\beta \in \left[10,100\right]$ to force the disk into a gravitationally unstable state. We characterized the ensuing gravito-turbulence with a focus on the spiral patterns commonly observed in self-gravitating disks. Our main findings may be summarized as follows.
\begin{enumerate}
\item Nonlocal energy transport is negligible in all our simulations: the conversion from orbital to thermal energy happens at the same radius. The angular momentum flux is dominated by the gravitational stress and it matches the value expected for viscous $\alpha$ disks in thermal equilibrium upon averaging.
\item Gravito-turbulence generates spiral density wakes that intermittently form and vanish over orbital timescales. These wakes are corotating with the gas at every radius to a good approximation. Large-scale spiral arms only manifest transiently through the coalescence of several neighboring wakes which then are sheared apart. 
\item The spiral wakes result from nearly symmetric compressive motions of supersonic amplitude in the midplane, making them the main site of energy dissipation. Simultaneously, these wakes promote vertical mixing by lifting the gas off from the midplane over several disk scale heights.
\item At every radius, the prominent spiral wakes are restricted to a narrow range of pitch angles which only weakly depend on the disk mass and cooling time. We can thus predict the morphology of large-scale apparent self-gravitating spirals for given disk rotation and mass surface density profiles.
\end{enumerate}

The numerical experiments presented in this paper cover a limited parameter space with a prescribed disk structure and simplified thermodynamics. In particular, our self-similar disk model is designed to promote locality in GI turbulence, and to reduce global effects as might be generated by edges or radial substructures. Circumstellar disks are known to exhibit substructures \citep{dsharp2} and to be gravitationally stable in their inner regions due to stellar irradiation \citep{alessio98}. Such radial structure would certainly affect the excitation and propagation of waves in massive disks \citep{lin64,papaloizou91,linpap11}. Global waves may also result from introducing a characteristic tidal forcing, as is the case in circumbinary disks for example \citep{marzari09,desai19}. 

Concerning the gas thermodynamics, we excluded the $\beta \lesssim 3$ regime of disk fragmentation to keep the disk relatively smooth and well-behaved. While disk fragmentation is an avenue of research on its own, it is also unclear how a more detailed radiative energy transport and the thermal stratification of the disk affect the spectrum of unstable perturbations and their nonlinear saturation in global geometry \citep{tsukamoto15,hirose17,hirose19}. 

Finally, the vertical flows accompanying spiral wakes can entrain dust grains away from the disk midplane \citep{riols20}. GI turbulence may thus affect the vertical segregation of different dust sizes \citep{pinte16,villenave20} or induce local enhancements of the dust disk thickness \citep{benisty15,dong15}. These 3D flows also open a new route of magnetic field amplification via a spiral dynamo in sufficiently ionized disks \citep{riols19, deng20}. 

\begin{acknowledgements}
  We thank Richard P. Nelson for his feedback on our preliminary results.
  We thank our referee, Doug Lin, for his questions and comments that helped clarify the final manuscript. 
  William B\'ethune acknowledges funding by the Deutsche Forschungsgemeinschaft (DFG) through Grant KL 650/31-1.
  The authors acknowledge support by the High Performance and Cloud Computing Group at the Zentrum f\"ur Datenverarbeitung of the University of T\"ubingen, the state of Baden-W\"urttemberg through bwHPC and the DFG through grant no INST 37/935-1 FUGG.

\end{acknowledgements}

\bibliographystyle{aa}
\bibliography{biblio}

\begin{appendix} 

  \section{Self-gravity in \textsc{Pluto}} \label{app:poisson}

  \subsection{Poisson solver}
  
  To determine the gravitational potential of the gas we solved Poisson's equation \eqref{eqn:poisson} with a method similar to that presented in Appendix B of \citet{bethune19}. This equation is cast into a matrix inversion problem after spatial discretization. We used the IBiCGStab algorithm \citep{yang02} implemented in the \textsc{PETSc} library \citep{balay97,balay19} to handle the parallel inversion of the Laplacian matrix. The convergence of the solver was characterized by a tolerance of $10^{-9}$ on the relative decrease of the residual norm between successive iterations. 

  The Laplacian operator was discretized by second-order finite differences, accounting for the nonuniform spacing and spherical geometry of the grid. Boundary conditions were applied in the rows corresponding to boundary cells in the matrix and right-hand-side vector. We used periodic boundary conditions in the azimuthal direction and Dirichlet (imposed $\Phi_{\mathrm{disk}}$ value, see below) on all the other boundaries. The Poisson problem was solved once per hydrodynamic timestep. Although the time-integration scheme formally becomes first-order accurate, \citet{bethune19} showed that it is able to accurately reproduce the growth rates of Jeans instability.

  \subsection{Implementation of boundary conditions}
  
  In its differential form, the Poisson problem allows any choice of values for $\Phi_{\mathrm{disk}}$ on the $\left(r,\theta\right)$ boundaries. In principle, one could impose $\Phi_{\mathrm{disk}}$ so as to describe the gravitational influence of massive bodies external to the computational domain. However, there is a single boundary distribution of $\Phi_{\mathrm{disk}}$ corresponding to a disk isolated in vacuum. This distribution is straightforward to obtain from the integral from of Poisson's equation:
  \begin{equation} \label{eqn:poissonint}
    \Phi\left(\bm{r}\right) = - G \int \frac{\rho\left(\bm{r'}\right)}{\vrac{ \bm{r} - \bm{r'} }} \dd^3 \bm{r'},
  \end{equation}
  where the integral spans the entire computational domain. While this integral can easily be implemented, it can also become prohibitively expensive to evaluate on all boundary cells. 
  
  We followed the method described by \citet{mueller95} to evaluate the integral \eqref{eqn:poissonint} at a reduced computational cost for a moderate accuracy loss. The Green's function $1 / \vrac{\bm{r} - \bm{r'}}$ is expanded onto spherical harmonics $Y_l^m$ and the series is truncated at a finite order $l\leq L$ much smaller than the number of cells in $\varphi$. After storing some integration weights, the algorithm reduces to two steps per Poisson problem. First, we compute the low-order moments of the density distribution by projecting it onto the $Y_l^m$. Then, we evaluate \eqref{eqn:poissonint} as a finite sum over these low-order moments. 

  We used the recursion relations highlighted by \citet{mueller95} to further reduce the number of operations involved. For subdomains of size $n_r\times n_{\theta} \times n_{\varphi}$, parallel communications amount to two reductions (sums) of $2 n_r\times\left(L+1\right)^2$ values in the $\left(\theta,\varphi\right)$ dimensions and two sums of $2 n_p \times \left(L+1\right)^2$ values across the $n_p$ subdomains in the radial dimension. At the heart of this method, the spherical harmonics were evaluated via the stable recursion relation provided by \citet{press07}. 

  Although this algoritm makes it computationally affordable to evaluate \eqref{eqn:poissonint} on the domain boundaries, the number of operations involved with the management of large arrays restrict the maximum order $L$ of the harmonic expansion. Luckily, this expansion quickly converges with order $l$ for smooth density distributions. As a compromise between speed and accuracy, we found that a truncation order $L=4$ was sufficient to capture the main features of the gravitational potential on the domain boundaries. 
  
  \subsection{Tests}

  We designed two experiments to test the correct implementation of the boundary conditions using spherical harmonics. In the first experiment, we meshed the spherical domain $\left(r,\theta,\varphi\right) \in \left[1,32\right] \times \left[0,\uppi\right] \times \left[0,2\uppi\right]$ with $512 \times 32 \times 64$ grid cells, with a logarithmic spacing in $r$ and uniform spacings in $\theta$ and $\varphi$. We uniformly distributed a mass $M=1$ inside the spherical shell $2 \leq r \leq 4$ and kept the density $\rho = 10^{-16}$ outside. The gravitational potential of the gas should then be constant inside $r\leq 2$ and vary as $-GM/r$ outside $r>4$. The integral \eqref{eqn:poissonint} was evaluated on both radial boundaries while the $\theta$ and $\varphi$ boundary conditions respected the spherical topology of the domain. Even though the monopole $L=0$ completely describes this density distribution, we expanded the series to $L=4$ to confirm that multipolar orders do not manifest themselves in this situation. 

  We obtained a spherically symmetric potential $\Phi_{\mathrm{disk}}\left(r\right)$ as expected. Outside $r>4$, the potential followed the expected $-GM/r$ to better than $5\times 10^{-3}$ accuracy all the way to the outer boundary. Inside $1 \leq r\leq 2$ the gravitational potential varied by less than $2\times 10^{-4}$ of its boundary value. At the inner radius $r\lin=1$ the potential matched the expected value $\Phi_{\mathrm{disk}}\left(r\lin\right) \approx -0.321 GM/r\lin$ to $4\times 10^{-4}$ acccuracy. These residual errors mainly come from the mass shell not being exactly bounded by $2\leq r \leq 4$ on the radial grid. We can thus confirm that the boundary potential satisfies Poisson's equation in its integral form \eqref{eqn:poissonint}, with the correct prefactors and spatial scalings. 

  \begin{figure}
    \includegraphics[width=1.0\columnwidth]{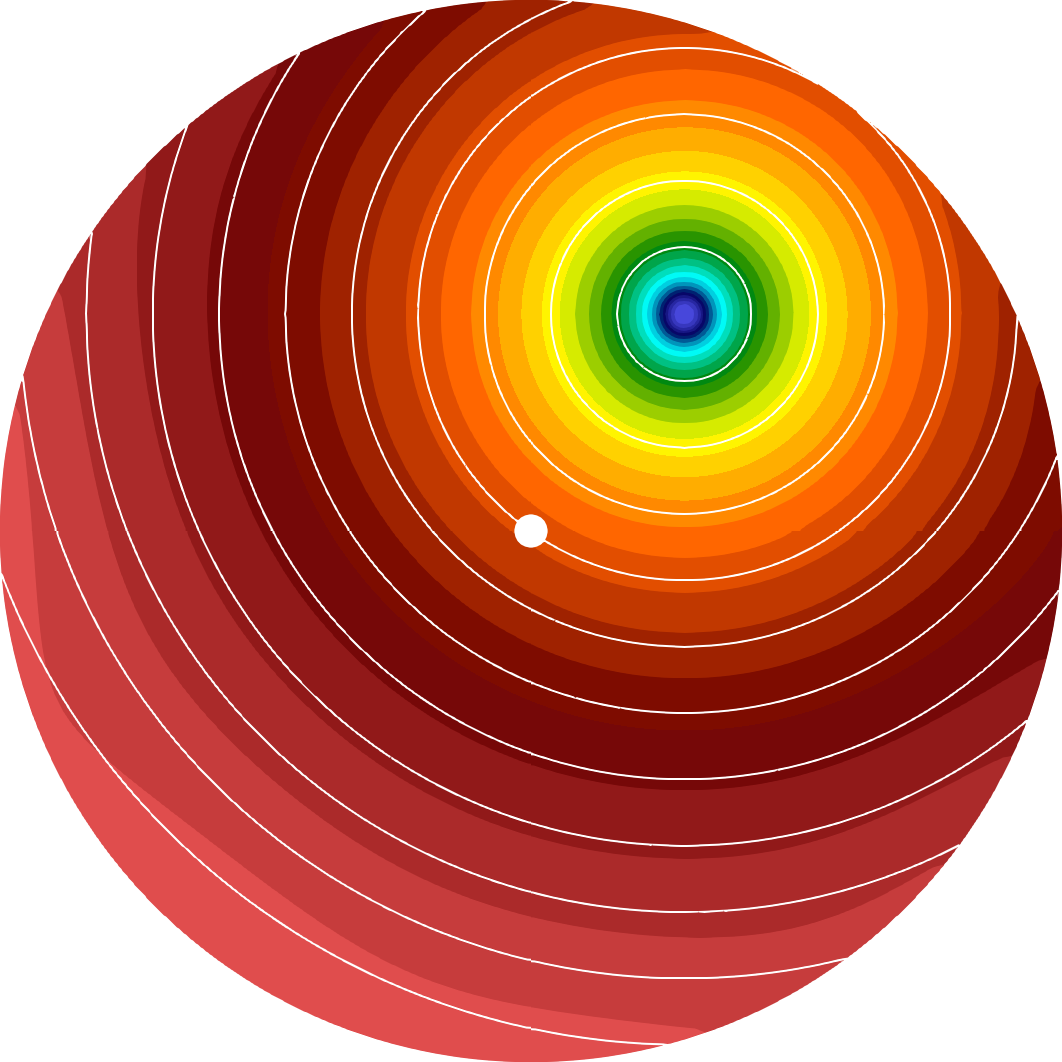}
    \caption{Gravitational potential (color-filled contours) of a massive ball inside a spherical domain. The domain extends radially over $r\in \left[1,32\right]$, the massive ball has a radius of $1$ and is offset from the origin by a cartesian displacement $x=y=z=16/\sqrt{3}$. This slice intersects the origin and is normal to the vector $\left(0,-1,1\right)$ so as to intersect the mass. The boundary values of the potential are evaluated using spherical harmonics of order up to $L=8$. The white circles are centered on the mass and correspond to theoretical iso-potential contours. The white disk $r<1$ at the center is excluded from the computational domain.}
    \label{fig:phithick}
  \end{figure}
  
  In the second experiment, we tested the convergence of the boundary conditions with the truncation order $L$ in an asymmetric configuration. We used the same domain and boundary conditions as in the first test, but we increased the resolution to $256 \times 192 \times 384$ so the grid cells are roughly cubic. We distributed a mass $M=1$ uniformly within a sphere of radius $1$ and center $x=y=z=16/\sqrt{3}$ while keeping $\rho = 10^{-16}$ outside. With $L=8$ the two arrays of integration weights already amount to $$ \underbrace{2}_\text{complex}\times \underbrace{256 \times 192 \times 384}_\text{grid} \times \underbrace{\left(8+1\right)^2}_\text{moments} \times \underbrace{8 \,\mathrm{bytes}}_\text{double} \approx 24\mathrm{Gb}$$ each --- in total, to be subdivided per process --- when stored in double precision.

  We show in Fig. \ref{fig:phithick} a slice of the sphere intersecting both the origin and the massive ball in this test with $L=8$. The key point is that the iso-potential contours remain circular when crossing the radial boundaries. This property is accurately satisfied in the vicinity of the mass: the gravitational potential behaves as if there were no boundaries at all, as intended. Deviations from circular contours only appear in the lower-left corner of Fig. \ref{fig:phithick}. These are expected because the density distribution is all but smooth in this test, leading to a slow convergence of the harmonic series with $L$. These two tests confirm that the boundary conditions are properly implemented and that subsequent errors would primarily come from the finite truncation order $L$.

\section{Reaction of the central object} \label{app:inertia}

In this paper, we considered disks orbiting a point-like massive object located at the origin of the frame of reference. By assuming that this frame was inertial, we neglected the acceleration of the central object due to the attraction of massive inhomogeneities in the disk. To test the validity of this approximation, we ran four additional simulations including noninertial effects. We kept the same simulation setup as before apart from dividing the grid resolution by two in every dimension for the sake of computational time. At every timestep we computed the acceleration vector
\begin{equation} \label{eqn:axelstar}
\bm{a}_{\star} = \sum_{i,j,k} \frac{\left[ G \,\rho \,\delta V \,\bm{r} \right]_{i,j,k}}{r_{i,j,k}^3}
\end{equation}
of the central object by summing over every grid cell indices $\left(i,j,k\right)$ in the computational domain. For the central object to remain at the origin of the frame, the disk must feel a uniform and opposite acceleration $-\bm{a}_{\star}$ which we included as a source term in the momentum equation. We fixed the dimensionless cooling time $\beta=10$ and sampled disk masses $M \in \left\lbrace 1/5, 1/3, 1/2, 1 \right\rbrace$.

\begin{figure}
  \includegraphics[width=1.0\columnwidth]{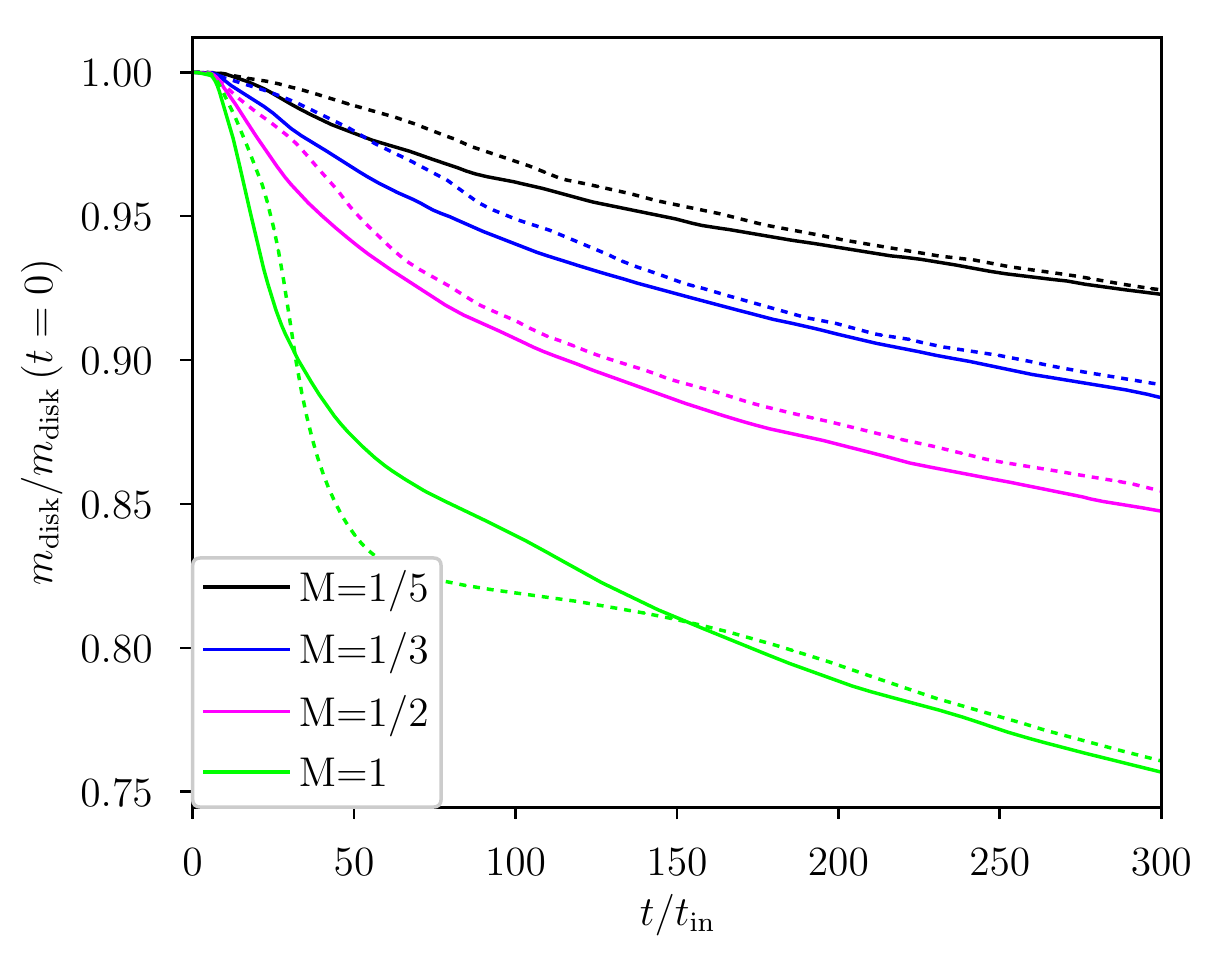}
  \caption{Total mass contained inside the computational domain relative to the initial mass as a function of time for different initial disk masses (see legend), with the central object reacting to the disk gravity (solid lines) and without (dashed lines).}
  \label{fig:stardisk_m}
\end{figure}

Because mass can flow out of the computational domain, it is not possible to track the exact position of the barycenter over time. Asymmetric accretion bursts can cause a momentary acceleration of the central object in one direction, with no simple means to guarantee global momentum conservation after the gas has left the computational domain. We found that such asymmetric accretion bursts mainly occur during the onset of gravito-turbulence. As shown in Fig. \ref{fig:stardisk_m}, the total mass losses amount to almost $25$ per cent over $300 t\lin$ in the most massive $M=1$ case. While the net momentum lost tends to vanish upon time averaging, it is generally different from zero and introduces a bias in the acceleration \eqref{eqn:axelstar} which we cannot simply correct for. 

\begin{table}
\caption{Simulations including the reaction of the central object: initial disk mass, mean dimensionless stress, mean spiral pitch angle and standard deviation of the gas surface density.}
\label{tab:reaction}
\centering          
\begin{tabular}{l c c c}
\hline\hline
$M$ & $\alpha$ & $\tan \left( i \right)$ & $\std{\Sigma}/\Sigma$\\
\hline
$1$ & $\left(1.0\pm 0.1\right)\times 10^{-1}$ & $0.29 \pm 0.04$ & $0.62 \pm 0.03$\\
$1/2$ & $\left(1.1\pm 0.1\right)\times 10^{-1}$ & $0.25 \pm 0.03$ & $0.62 \pm 0.01$\\
$1/3$ & $\left(1.0\pm 0.1\right) \times 10^{-1}$ & $0.26 \pm 0.02$ & $0.68 \pm 0.03$\\
$1/5$ & $\left( 8.3 \pm 0.9 \right) \times 10^{-2}$ & $0.21 \pm 0.01$ & $0.65 \pm 0.02$\\
\hline                  
\end{tabular}
\end{table}

Once in a steady gravito-turbulent phase, the disk remains Keplerian and develops no substantial eccentricity. Compared to the inertial simulations, the diagnostics gathered in Table \ref{tab:reaction} are overall compatible with those of Table \ref{tab:recap} for a given disk mass. The slight differences found for the least massive $M=1/5$ case likely come from the reduced grid resolution. The accretion histories drawn in Fig. \ref{fig:stardisk_m} are also similar after the onset of turbulence $t\gtrsim 100 t\lin$. We conclude that the reaction of the central object does not significantly alter our diagnostics regarding the disk dynamics, in agreement with \citet{michael10}. This conclusion is also supported by the simulations of \citet{forgan11}, who found that $m=1$ perturbations remain subdominant even for disk masses $M=3/2$ of the central mass. The stability analysis of \cite{heemskerk92} predicts an $m=1$ instability only when $M \gtrsim 1$, and it does not account for the effective viscosity of the gravito-turbulent disk, which may explain why our results with and without the reaction of the central object are compatible.

\begin{figure}
    \includegraphics[width=1.0\columnwidth]{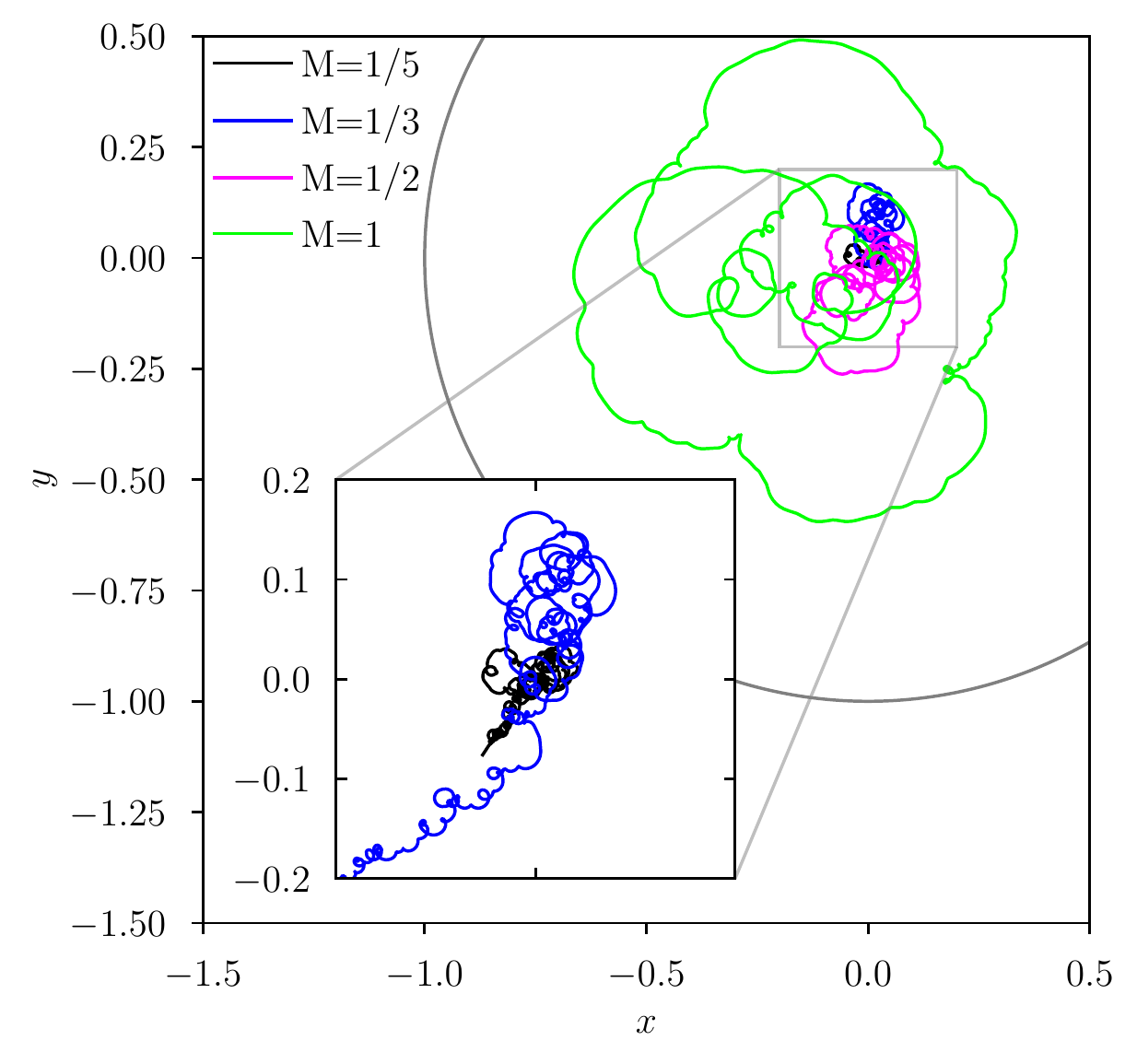}
    \caption{Trajectory of the central object reacting to the disk gravity from $100 t\lin$ to $300 t\lin$ in the equatorial plane for different disk masses (see legend). The gray arcs represent the inner boundary of the computational domain at a radius $r=1$. The inset zooms on the innermost regions for the two least massive disks.}
    \label{fig:stardisk_xy}
  \end{figure}

By integrating the individual components of the acceleration vector $\bm{a}_{\star}$ over time, one can reconstruct the hypothetical trajectory of the central object around the barycenter. Since the exact position of the barycenter cannot be known, this trajectory only serves to estimate the typical displacements of the central object. Reversing the sign of the acceleration vector, we equivalently obtain the typical displacements of the disk. We begin the integration at $100 t\lin$ to discard the initial transient phase and show the resuling trajectories in Fig. \ref{fig:stardisk_xy}. All the trajectories remain confined inside $r/r\lin<1$, so the disk is never significantly displaced through the inner radial boundary. As expected, the displacements are increasingly large for larger disk masses, reaching up to $r \approx 0.6 r\lin$ in the $M=1$ case. The trajectories of the two most massive cases $M\geq 1/2$ roughly describe orbits but with no clear cyclic pattern, as one would expect if the disk supported long-lived $m=1$ density perturbations.
  
\end{appendix}

\end{document}